\documentclass[journal abbreviation, manuscript]{copernicus}

\usepackage[utf8]{inputenc}
\usepackage[T1]{fontenc}
\begin{document}

\nolinenumbers

\title{Statistical noise and missing forcing limit estimates of Earth's feedback from prescribed sea-surface temperature simulations}




%







\Author[1][gergana.gyuleva@env.ethz.ch]{Gergana}{Gyuleva}
\Author[1]{Reto}{Knutti}
\Author[1]{Robin}{Noyelle}
\Author[1]{Urs}{Beyerle}
\Author[1]{Erich}{Fischer}
\Author[2]{Sebastian}{Sippel}

\affil[1]{Institute for Atmosphere and Climate Science ETH Zurich, Zurich, Switzerland}
\affil[2]{Institute for Meteorology, Leipzig University, Leipzig, Germany}




\runningtitle{TEXT}

\runningauthor{TEXT}

\received{}
\pubdiscuss{} 
\revised{}
\accepted{}
\published{}


\firstpage{1}

\maketitle

\begin{abstract}  \newline
Earth's feedback parameter measures how the Earth system responds to forcing and is inversely proportional to climate sensitivity. Sea-surface temperature (SST) patterns can modulate the value of the feedback parameter. Differences between observed and simulated SSTs have raised the question how the observed SSTs evolution impacts the global feedback. The standard method for estimating this effect uses observed SSTs prescribed to an atmospheric model with fixed pre-industrial atmospheric forcing. This method makes two assumptions: first, that the observed SSTs capture all relevant effects from the forcing, so that prescribing a time-varying forcing is unnecessary; second, that the temporal variations in the feedback parameter are driven by the evolving SST pattern and can be estimated via moving-window regressions. We test these assumptions by running controlled experiments in which SSTs from fully-coupled historical simulations are prescribed to an atmospheric model. We find that the prescribed-SST experiments fail to capture the coupled feedback evolution. This is explained by two effects: First, the absence of prescribed atmospheric forcing, and second, statistical noise arising from the computation of moving-window regressions. We find no evidence of any significant relationship between evolving SST patterns and changes in the feedback time series in a 4000-year pre-industrial control simulation. Any trends in the feedback parameter detected on timescales shorter than $\sim$100 years are indistinguishable from statistical noise, making their attribution to the evolving SST pattern extremely difficult. Our results imply that prescribed SST simulations offer limited potential for inferring temporal changes in Earth's parameter over the observational period.

\end{abstract}


\introduction  
\subsection{The climate feedback parameter and its relation to future warming}
Understanding and quantifying the sensitivity of Earth's climate to anthropogenic greenhouse gas (GHG) emissions is one of climate science's most fundamental questions. The Intergovernmental Panel on Climate Change's (IPCC) 6\textsuperscript{th} assessment report (AR6) estimates a "likely" range of 2.5--4~K for the equilibrium temperature after a CO\textsubscript{2}-doubling, known as Equilibrium Climate Sensitivity (ECS). This assessment is based on combined evidence from climate feedbacks estimated from current observations and general circulation models (GCMs), historical trends in the global energy budget, and paleo-climatic evidence \citep{forster_earths_2021,sherwood_assessment_2020}. In 1979, half a century earlier, the Charney Report provided a likely range of 1.5--4.5~K for ECS based on just two early-generation GCMs. This raises the question: why has the uncertainty in future warming remained so persistently, and where does it come from? \newline

One can view the large uncertainty in future warming through the lens of a forcing-feedback framework \citep[e.g.,][]{gregory_new_2004}: the energy imbalance $N$ of the Earth system---the difference between the outgoing long-wave (LW) radiation and net (incoming minus reflected) short-wave (SW) solar radiation at the top-of-atmosphere (TOA)---can be modeled as 
\begin{equation}
    N = R + F.
\end{equation}
The forcing-feedback framework assumes that $N$, which corresponds to the heat uptake of the system (in watts per square meter), can be separated into a radiative forcing term $F$ and a response term $R$. Importantly, the forcing term $F$ due to changes in the atmospheric composition is independent of the system state, while $R$ is typically modeled as being (linearly) dependent on the temperature $T$ of the system \citep{gregory_new_2004,north_energy_1981,wigley_analytical_1985,budyko_effect_1969,sellers_global_1969}:
\begin{equation}
    R = R(\Delta T)\simeq \lambda \Delta T.
\end{equation}
The $\Delta T$ represents the change in the system's temperature as a result of the imposed forcing $F$, while $\lambda$ is the linear feedback parameter:
\begin{equation}\label{eq:dR/dT}
    \lambda = \frac{\partial R}{\partial T}
\end{equation}
One can then write  
\begin{equation}\label{eq:linear_fff}
    N = \lambda \Delta T + F.
\end{equation}
The feedback relates a given temperature change $\Delta T$ induced by the forcing $F$ to the radiative response $R$, and is negative for a stable climate. It can therefore be interpreted as the restoring force of the system after a perturbation imposed by $F$ \citep{roe_why_2007}. \newline 


The importance of the feedback parameter lies in its relation to ECS: for a forcing $F_{\textrm{2xCO\textsubscript{2}}}$ resulting from a doubling of the atmospheric CO\textsubscript{2} concentration, ECS is given by the temperature difference $\Delta T_{\mathrm{eq}} $ when the system reaches radiative equilibrium ($N = 0$), resulting in 
\begin{equation}
    \textrm{ECS} := \Delta T_{\mathrm{eq}} = - \frac{F_{\textrm{2xCO2}}}{\lambda}
\end{equation}
The feedback parameter therefore directly provides information about how sensitive Earth's climate is to perturbations: a strongly negative feedback means Earth is able to restore equilibrium after a forcing perturbation---such as an increase in CO\textsubscript{2} concentration---with comparatively little increase in temperature, i.e., the system is efficient in restoring a balance resulting from the extra heat accumulated due to the forcing. A feedback parameter closer to zero (less negative), on the other hand, implies a system which needs to heat up more in order to restore radiative equilibrium. Due to its inverse proportionality to ECS, the feedback parameter is particularly relevant in determining future warming \citep{sherwood_assessment_2020}. \newline

\subsection{Existing methods to quantify the feedback parameter}
There have been extensive efforts to quantify the feedback (i.e., ECS) in both models and observations. This endeavor is not straightforward, for multiple reasons. First, the linear feedback framework in Eq.~\ref{eq:linear_fff} assumes a constant value for $\lambda$ that is independent of timescale and type of forcing. Under this assumption, estimating $\lambda$ can be done via the so-called `Gregory plot' \citep{gregory_new_2004}: $\lambda$ is retrieved as the slope of the regression of $N$ onto $\Delta T$ over the first 100-200 years in abrupt-CO\textsubscript{2}-quadrupling experiments. The linearity assumption of $\lambda$ being constant is, however, only an approximation \citep{knutti_feedbacks_2015}, and in reality $\lambda$ has been shown to be forcing-dependent \citep[e.g.,][]{marvel_implications_2016,yoshimori_dependency_2011}, state-dependent \citep[e.g.,][]{boer_climate_2005,meraner_robust_2013}, and to vary due to evolving sea-surface temperature (SST) patterns via the so-called pattern effect \citep[e.g.,][]{senior_time-dependence_2000,armour_time-varying_2013,andrews_dependence_2015,rugenstein_dependence_2016,ceppi_relationship_2017}. Early work on the pattern effect demonstrated that evolving SST patterns emerging when letting the GCM equilibrate in an abrupt-CO\textsubscript{2}-increase experiment result in a curvature of the Gregory plot towards a weakening of the feedback $\lambda$ \citep{rugenstein_dependence_2016,andrews_dependence_2015}. Furthermore, several studies using linear response theory have shown that Earth's feedback is best approximated not by a single $\lambda$, but rather three distinct but constant $\lambda$'s, each governing a different component of the Earth system with its own characteristic timescale \citep[e.g.,][]{fredriksen_estimating_2021,fredriksen_21st_2023,proistosescu_slow_2017}. \newline

The pattern effect has gained special attention in recent years, as the SST patterns modeled by state-of-the-art coupled atmosphere-ocean GCMs (AOGCMs) fail to reproduce some of the observed trends in SSTs \citep{seager_strengthening_2019,olonscheck_broad_2020,wills_systematic_2022,watanabe_enhanced_2021}. Numerous studies suggest that the observed pattern of warming---with a strengthening of the East-West tropical Pacific SST gradient, for example---is one producing more negative feedbacks, similar to those during a La Ni\~{n}a phase \citep{wills_systematic_2022,ceppi_relationship_2017,armour_sea-surface_2024}. In particular, idealized studies based on Green's functions have demonstrated that cloud and lapse-rate feedbacks in the tropical Pacific are particularly sensitive to the pattern of warming \citep{zhou_analyzing_2017,dong_attributing_2019}. This motivates the need to estimate the effect of the evolving observed SST pattern on the time-evolution of the global feedback parameter.\newline

The second major difficulty in quantifying a time-varying global feedback parameter from observations is the need for accurate and long observational time series of $N$, $F$, and $T$. However, uncertainties in estimates of $F$ are large, and satellite observations of $N$ exist only from 2001 onwards \citep{loeb_clouds_2018}, leading to highly uncertain observational estimates of the feedback \citep{meyssignac_time-variations_2023}. For this reason, the state-of-the-art method to estimate the observed feedback evolution employs atmospheric GCMs forced by observed SSTs and sea-ice concentration (SIC), introduced formally as \textit{amip-piForcing} runs by \citet{gregory_variation_2016} (although SSTs have been prescribed to atmospheric models in many other earlier works for different purposes, e.g., \citet{gates_overview_1999,boer_results_1992,cess_interpretation_1989,andrews_using_2014}). \newline

The idea behind the \textit{amip-piForcing} setup for estimating the feedback evolution is that by prescribing the observed SST and SIC pattern as boundary conditions to the atmospheric model, one captures all relevant effects for the historical feedback evolution. The \textit{amip-piForcing} method makes two important assumptions. First, atmospheric forcing is kept constant at preindustrial levels in \textit{amip-piForcing} simulations, allowing to set $F=0$ in Eq.~\ref{eq:linear_fff}. This step implicitly assumes that all feedback-relevant processes of the true historical forcing $F$ are sufficiently well-captured in the observed SST and SIC evolution used as boundary conditions \citep{gregory_variation_2016}. The assumption $F=0$ allows for a simplification of the feedback calculation by avoiding the need of cumbersome estimations of $F$. The feedback can now be estimated simply from time series of $\Delta T$ and $N$, which can be directly retrieved from the \textit{amip-piFrocing} run output.\newline 

The second assumption lies in the way the feedback is estimated from $N$ and $\Delta T$ in practice: since the goal of the \textit{amip-piForcing} runs is to quantify the effect of the evolving SST pattern on the evolving feedback $\lambda$, one wants to obtain a time series of $\lambda$ which should capture possible SST-driven trends. In practice, this is done via 30-year moving window regressions of $N$ against $\Delta T$, with $\lambda$ being the slope of the regression: 
\begin{equation}\label{eq:amip_pf_movwin}
    \lambda(t) = \left. \frac{\partial N\left(\tilde{t}\right)}{\partial T \left(\tilde{t}\right)} \right|_{\tilde{t} \,\in\, (t-30,\, t]}
\end{equation}
By doing so, the implicit assumption made is that time variations in $\lambda(t)$ computed in this way are physically interpretable and result from SST variability.\newline

Under these assumptions, numerous studies have used \textit{amip-piForcing} runs to quantify the feedback evolution of the real climate system based on prescribed SST patterns \citep{gregory_variation_2016,andrews_accounting_2018,zhou_greater_2021,andrews_effect_2022,dong_biased_2021}. These studies all come to the conclusion that the feedback shows a strong trend towards more negative, stabilizing values in recent decades and attribute it to the specific pattern of SSTs. In particular, \citet{dong_biased_2021} explicitly compare the observed feedback evolution as derived from \textit{amip-piForcing} runs via Eq.~\ref{eq:amip_pf_movwin} to the evolution of the feedback in coupled \textit{historical} runs from CMIP6 (computed as in Eq.~\ref{eq:amip_pf_movwin}, but with $\partial (N - F)$ in the denominator, using an $F$ estimate). They find that the \textit{historical} simulations do not reproduce the stabilizing trend of the \textit{amip-piForcing} simulations post-1980. They then attribute this effect to the difference in SST pattern between \textit{historical} simulations and observations. This finding has wide-reaching implications, as a weaker or stronger feedback directly relates to a different future warming. \newline 

Given the importance of the \textit{amip-piForcing} runs, it is crucial to test (I), whether they capture all relevant information to estimate the feedback when setting $F=0$; and (II), whether the variability found in $\lambda(t)$ using 30-year moving window regressions is driven by SST pattern variability. We test the first assumption by forcing the atmospheric component of the CESM2 model \citep{danabasoglu_community_2020} with evolving SSTs and SIC concentration in an \textit{amip-piForcing} setup, but with SSTs and SIC taken from coupled \textit{historical} simulations of CESM2 (Sec.~\ref{sec:results:1_F_hypothesis}). This allows for a direct comparison of the \textit{amip-piForcing} feedback to the feedback estimated in the corresponding coupled \textit{historical} run with identical SSTs and SIC boundary conditions. Our results uncover the inability of the \textit{amip-piForcing} method to recover the feedback of the corresponding \textit{historical} run with identical SSTs and SIC. We next show that this is in large part due to the omission of atmospheric forcing in the \textit{amip-piForcing} setup, thus challenging the first assumption underlying the \textit{amip-piForcing} setup. \newline

Next, we study sources of variability in the feedback time series of a coupled pre-industrial control (\textit{piControl}) simulation (Sec~\ref{sec:results_Variability}). Here, $F=0$ by definition and if the second main assumption holds, variability in $\lambda(t)$ should be driven by the SST pattern. We find that variability on timescales shorter than $\sim$100 years is primarily driven by an artificial statistical effect arising from applying a moving window regression over $N$ and $T$. Our findings imply that even if SSTs are driving variability in $\lambda$, the causal effect is difficult if not impossible to separate from the statistical variability generation of the moving window process. Furthermore, we find no evidence of a link between specific SST patterns and the feedback evolution. Our results imply that a) trends found in $\lambda(t)$ from \textit{amip-piForcing} runs cannot be clearly attributed to evolving SST patterns, and b) that any trends measured in $\lambda(t)$ on timescales shorter than $\sim$100 years are within the typical variability timescale of $\lambda$ and therefore potentially indistinguishable from internal variability. 

\section{Data \& Methods}
\subsection{Experimental setup}
We conduct atmosphere-only runs with the Community Atmosphere Model Version 6 (CAM6), which is the atmospheric component of the Community Earth System Model Version 2 (CESM2) \citep{danabasoglu_community_2020}. We conduct three distinct types of experiments with CAM6: \textit{amip-piForcing}-style runs and \textit{amip-histForcing}-style runs are outlined below and form the basis for our analysis. Additionally, we perform \textit{piClim}-style runs required for estimating the time-varying forcing $F$ \citep{pincus_radiative_2016}, which are detailed in Appendix~\ref{sec:app:meth_diagnosing_F}. The individual simulations are summarized in Table~\ref{tab:simulation_data}. 

\subsubsection*{``amip-piForcing''-style runs}
These simulations are identical to the \textit{amip-piForcing} setup \citep{gregory_variation_2016,webb_cloud_2017}, except having different (i.e., simulated rather than observed) SST and SIC boundary conditions. In all \textit{amip-piForcing}-style simulations we fix atmospheric forcing at preindustrial (1850) levels. We call these runs \textit{amip-piForcing}-[...] in Table~\ref{tab:simulation_data}. The [...] stands for the different SST and SIC used as boundary conditions. We conduct three distinct types of these \textit{amip-piForcing}-style experiments, differing only in the prescribed SST and SIC boundary conditions:
\begin{itemize}
    \item \textit{amip-piForcing-histSST}: For these 11 runs we prescribe SST and SIC from from 11 fully coupled \textit{historical} initial condition members of the CESM2 model performed as part of CMIP6 \citep{danabasoglu_community_2020}. The forcing in the coupled \textit{historical} runs is prescribed according to the CMIP6 protocol with historically evolving best-estimates of all known forcing agents from 1850--2014 \citep{eyring_overview_2016}. The SSTs taken from those runs therefore exhibit a warming over the historical period.
    \item \textit{amip-piForcing-ghgSST}: For these 11 runs we prescribe SST and SIC from 11 fully coupled initial condition members from the CESM2 single-forcing large ensemble which have historically evolving GHG-concentrations as forcing \citep{simpson_cesm2_2023}. The coupled runs from which we take the boundary conditions are called \textit{GHG2}, following the convention of \citet{simpson_cesm2_2023}. The coupled \textit{GHG2} runs therefore have all other forcings fixed at 1850 levels, apart from GHG forcing, which evolves as in the \textit{historical} simulation \citep{simpson_cesm2_2023}. The SSTs of these simulations also exhibit the GHG-induced warming over the historical period. 
    \item \textit{amip-piForcing-piSST}: For this run we prescribe SSTs and SIC from the fully coupled CESM2 \textit{piControl} simulation produced as part of CMIP6 \citep{danabasoglu_community_2020}. Forcing in the coupled \textit{piControl} simulation is held fixed at 1850 levels \citep{danabasoglu_community_2020}. We run the \textit{amip-piForcing-piSST} simulation for 410 years.
\end{itemize}
\begin{table}[htbp]
    \centering
    \scriptsize 
    \caption{AMIP-style simulations performed with the CAM6 model. \textit{piClim}-style runs are detailed in Appendix~\ref{sec:app:meth_diagnosing_F}.}
    \label{tab:simulation_data}
    \renewcommand{\arraystretch}{1.3}
    
    \begin{tabular}{
        >{\raggedright\arraybackslash}p{2.5cm} 
        >{\centering\arraybackslash}p{1.3cm} 
        >{\centering\arraybackslash}p{1.6cm} 
        >{\centering\arraybackslash}p{1.8cm} 
        >{\raggedright\arraybackslash}p{5.5cm}
    } 
        \hline 
        Name & Members & Simulation length / period & Forcing & SST \& SIC boundary conditions\\
        \hline 
        \textit{amip-piForcing-histSST}   & 11 & 1850--2014 & 1850 (fixed) & CESM2 historical SSTs and SICs from members r\textbf{X}i1p1f1, $ \mathrm{\textbf{X}}\in \{1,...,11\}$ in the CMIP6 ensemble (\cite{danabasoglu_community_2020}) \\ 
        \textit{amip-piForcing-ghgSST}   & 11 & 1850--2014 & 1850 (fixed) & CESM2 GHG2 SSTs and SIC from members 1--11 of the CESM2 single-forcing large ensemble (\cite{simpson_cesm2_2023})\\ 
        \textit{amip-piForcing-piSST}    & 1  & 405 years & 1850 (fixed) & CESM2 piControl SSTs and SIC (\cite{danabasoglu_community_2020})\\ 
        \textit{amip-histForcing-histSST} & 5  & 1850--2014 & historical   & CESM2 historical SSTs and SICs from members r\textbf{X}i1p1f1, $ \mathrm{\textbf{X}}\in \{1,...,11\}$ in the CMIP6 ensemble (\cite{danabasoglu_community_2020}) \\ 
        \textit{piClim-control}           & 1  & 200 years   & 1850 (fixed) & CESM2 piControl SSTs and SICs (climatology) (\cite{danabasoglu_community_2020})\\ 
        \textit{piClim-histall}           & 1  & 1850--2014 & historical   & CESM2 piControl  SSTs and SICs (climatology) (\cite{danabasoglu_community_2020})\\
        \hline 
    \end{tabular}
\end{table}
\subsubsection*{``amip-histForcing''-style runs}
These runs differ from the standard \textit{amip-piForcing}, because apart from prescribing time-varying SST and SIC, we also prescribe time-varying historical atmospheric forcing. One ensemble of such runs is performed (Table~\ref{tab:simulation_data}):
\begin{itemize}
    \item \textit{amip-histForcing-histSST}: These 11 runs have SSTs and SIC identical to the 11 members of the \textit{amip-piForcing-histSST} ensemble (and therefore, identical to the fully coupled \textit{historical} runs), but also time-varying historical forcing prescribed to the atmosphere. CAM6 in these runs therefore experiences an almost identical forcing from ocean and atmosphere as the corresponding 11 members from the coupled \textit{historical} CESM2 runs (up to variability from the initial conditions and numerical differences). 
\end{itemize}

\subsection{Calculation of the feedback parameter $\lambda$ from numerical simulations}
The feedback is generally estimated by Eq.~\ref{eq:dR/dT}: \citep[e.g.,][]{gregory_new_2004,gregory_variation_2016}:
\begin{equation*}
    \lambda  = \frac{\partial R}{\partial T} 
\end{equation*}
In general, the radiative response $R$ is given by $R = N - F$. \newline

Each experimental setup implies a radiative response, which is calculated according to the $N$ and $F$ present in the corresponding experimental setup. All \textit{amip-piForcing}-style experiments and the piControl simulation imply no forcing by definition, i.e., $F=0$ \citep{gregory_variation_2016}. In all other coupled and atmosphere-only runs with time-varying forcing, $F$ needs to be estimated separately (see Sec.~\ref{sec:app:meth_diagnosing_F}).\newline

In order to obtain time series of $\lambda$, we follow the moving-window-regression method used in literature \citep[e.g.,][]{gregory_variation_2016,andrews_accounting_2018,andrews_effect_2022,dong_biased_2021}. This consists of taking 30-year moving windows, for which the numerator $N - F$ is regressed onto the denominator $\Delta T$, as in Eq.~\ref{eq:amip_pf_movwin}:
\begin{equation}\label{eq:general_feedback}
    \lambda(t) = \frac{\partial R \left(\tilde{t}\right)}{\partial T \left(\tilde{t}\right)} = \left. \frac{\partial \left( N - F \right) \left(\tilde{t}\right)}{\partial T \left(\tilde{t}\right)} \right|_{\tilde{t} \,\in\, (t-30,\, t]}
\end{equation}
The time series of $\lambda$ then results from the time series of the slopes of these moving-window regressions. $N$ is calculated from the CMIP6 variables as the difference in TOA downward shortwave and upward short-and longwave fluxes, \textit{rsdt - rsut - rlut}. For the simulations done with CESM2, we calculate the net TOA imbalance from the CESM2 output variables \textit{FSNTOA - FLUT}, being the difference of net shortwave and outgoing longwave radiation. For $T$ we use the variable \textit{tas} for CMIP6 output and the variable \textit{TREFHT} for CESM2 raw output. \newline

We estimate the forcing $F$ using three different methods, in order to test the sensitivity of our results to the $F$ estimate. These are outlined in detail in Appendix~\ref{sec:app:meth_diagnosing_F}. For all figures of the main text, we display results where $F$ has been estimated following the Radiative Forcing Model Intercomparison Project (RFMIP) method \citep{pincus_radiative_2016}. This method involves subtracting an atmosphere-only control run with fixed climatological SSTs, from an atmosphere-only run where the time-varying forcing is prescribed but SSTs are also held climatologically fixed (see Appendix~\ref{sec:app:meth_diagnosing_F}).

\subsection{Monte-Carlo simulations for variability analysis}\label{sec:meth:montecarlo}

In Sec.~\ref{sec:results_Variability} we turn to idealized coupled \textit{piControl} and atmosphere-only \textit{piClim-control} simulations in order to analyze variability in $\lambda(t)$. For this purpose, we attempt to model $T$ and $N$ from \textit{piControl} and \textit{piClim-control} statistically: Our goal is to find a good statistical model of $T$ and $N$ in these simulations, so that we can generate Monte-Carlo time series of $T$ and $N$ with similar statistical properties as the true time series.
\subsubsection*{Modeling $T$ and $N$ in \textit{piClim-control} statistically}
We hypothesize that $T$ and $N$ in the \textit{piClim-control} simulation can be modeled as jointly normally distributed. This is based on the fact that \textit{piClim-control} is forced by climatologically fixed, monthly-varying SSTs and SIC. We therefore expect to have no SST-driven variability in $N$ and $T$ on annual timescales. Since ocean memory is erased, we also do not expect to have any autocorrelation in $N$, $T$, or jointly in $N$ and $T$. To test the normality hypothesis, we perform first a uni-variate normality test for $N$ and $T$ separately based on \citet{dagostino_tests_1973}. We find p-values of 0.99 and 0.82 for $N$ and $T$, respectively, meaning that there is insufficient evidence to reject the null-hypothesis that $N$ and $T$ are normal. Similarly, the p-value for the multivariate-normality test based on \citet{henze_class_1990} is 0.49, meaning that we can justify the assumption of multivariate normality of $N$ and $T$.\newline

In order to generate Monte-Carlo simulations of $N$ and $T$ with the statistical properties of \textit{piClim-control}, we generate one million samples from a bi-variate Gaussian distribution with mean vector and covariance matrix estimated from the sample statistics of $N$ and $T$.

\subsubsection*{Modeling $T$ and $N$ in \textit{piControl} statistically}
In the coupled \textit{piControl} simulation we expect memory from ocean variability to lead to strong autocorrelation in $T$ \citep[e.g.,][]{hasselmann_stochastic_1976,frankignoul_stochastic_1977}. Since coupling between $N$ and $T$ is now free, we also expect there to possibly be cross-(auto)correlations between $N$ and $T$. For this reason we choose to model the \textit{piControl} time series of $N$ and $T$ via a vector autoregressive model (VAR) of order $k$ \citep{lutkepohl_new_2005}: Let $Y_t = \left(\begin{array}{@{}c@{}}
    T_{t} \\
    N_{t} 
    \end{array} \right)$, then we can model $Y_t$ via

\begin{equation}
    Y_t = \mu + A_1 Y_{t-1} + ... + A_k Y_{t-k} + u_t, ~~~~ \mathrm{where}~~~ u \sim \mathcal{N}\left(0, \Sigma_u\right) 
\end{equation}
The matrices $A_1, ..., A_k$ contain autoregression coefficients for lags $1,...,k$, and the noise term $u_t$ comes from a bivariate normal distribution with zero mean. We fit the VAR process with automatic optimal lag-order selection via the Akaike information criterion. This leads to an optimal lag order of $k=5$. We then generate one million years of $N$ and $T$ from the VAR(5) model fitted on \textit{piControl} data. \newline 

Spectral densities of the time series were computed with the multi-tapering method \citep{huybers_links_2006}. The computation was done over the whole length of the respective simulations, after linearly de-trending the time series of the variables $N$, $T$, and $\lambda$. 
\newline

\subsection{Bootstrapping of periods in \textit{piControl} for variability analysis}\label{sec:meth:bootstrap}

In Section~\ref{sec:res:bootstrap_piControl} we investigate whether periods of weakening or strengthening feedback can be associated to particular SST patterns. For this analysis, we use a 4000-year long fully-coupled pre-industrial control simulation performed with CESM2, which we term \textit{piControl-long}. It has identical boundary conditions as the CMIP6-version of the CESM2 \textit{piControl} introduced earlier \citep{danabasoglu_community_2020}. \newline

We compute the feedback in the \textit{piControl-long} simulation from 30-year moving window regressions as in Eq.~\ref{eq:general_feedback}. We then seek periods during which the feedback weakens (i.e., increases towards zero), or strengthens (i.e., becomes more negative). To select such periods we first smooth the feedback time series using a Savitzky-Golay filter with window length 81 and polynomial order 4 \citep{savitzky_smoothing_1964}. We then select periods where the derivative of the smoothed time series is positive or negative for at least twenty consecutive years. In total, we identify 49 periods of feedback strengthening and 49 of feedback weakening in the 4000-year long \textit{piControl} simulation. The time series of the feedback in the \textit{piControl} simulation together with the selected periods of weakening and strengthening is displayed in Fig.~\ref{fig:app:composite_periods}. Our results remain largely unchanged if the minimum period length is shortened to 10 or increased to 30 years. \newline

We then compute the surface air temperature trend for each grid cell during each period of weakening and strengthening. We then average over all periods of strengthening or weakening feedback, in order to obtain the composite of the surface air temperature trends. We also compute the agreement fraction on the sign of the trend between the samples that make up the composite. The agreement fraction is the number of samples in the composite which belong to the majority group (i.e., if the majority of samples show a positive trend in a grid cell, then we count the number of samples which show this positive trend), divided by the total number of samples in the composite (here, 49 for both weakening and strengthening feedback periods). \newline 

To test the significance of the composite surface air temperature trends and their agreement, we employ bootstrapping on the 4000-year long time series. We sample a set of non-overlapping periods of the same length and number as the ones used to compute the composites, 10\,000 times in total. This gives us a null-distribution of randomly sampled periods in \textit{piControl-long} during which any trends in the composites occur due to random noise. We would expect the average agreement on the sign of trends between periods of any bootstrap sample to be close to 0.5 (half of sampled periods show a positive, and half a negative trend). For the significance of the trend in a given grid cell, we compute the two-sided p-value as 
\begin{equation}
    p = 2 \cdot \min \{q, 1-q\}
\end{equation}
where $q$ is the empirical quantile of the composite trend with respect to the bootstrapped distribution of composites. For the significance of the agreement fraction we compute the one-sided p-value based on the empirical quantile $q$ of the agreement between the samples of the composite as 
\begin{equation}
    p = 1-q
\end{equation}
We use a one-sided p-value because the agreement fraction is $>=0.5$ by definition, and exactly $0.5$ when there is no agreement (i.e., when exactly half of the composites show a positive and the other half a negative trend). We therefore only test whether the agreement in our feedback strengthening or weakening composites is \textit{larger} than that of the bootstrapped distribution. \newline

With the above procedure, we perform a hypothesis test on each grid cell in order to test its significance, either for the trend of the composite, or for the agreement fraction among the composite. If the p-value for a given grid cell is smaller than our chosen significance level of $\alpha=5\%$, we reject the null hypothesis (the bootstrapped distribution) and term the values `significant', in the sense of being different from random noise. However, this procedure quickly leads to the problem of multiple testing, according to which we naturally expect to find around $\alpha$ of our samples to be significant, purely by construction of the statistical test \citep{livezey_statistical_1983,wilks_stippling_2016}. The problem is further aggravated by strong spatial correlations in climatic fields \citep{wilks_stippling_2016}. We therefore employ the correction by \citet{benjamini_controlling_1995} which controls for the false discovery rate at the 10\% level and is the recommended correction for use cases in climate science in the presence of spatial (auto)correlation \citep{wilks_stippling_2016}.

\section{Results}

\subsection{Comparing \textit{amip-piForcing}-style and coupled runs with identical SSTs \& SIC}\label{sec:results:1_F_hypothesis}
\begin{figure}[ht]
	\centering
	\includegraphics[width=\textwidth]{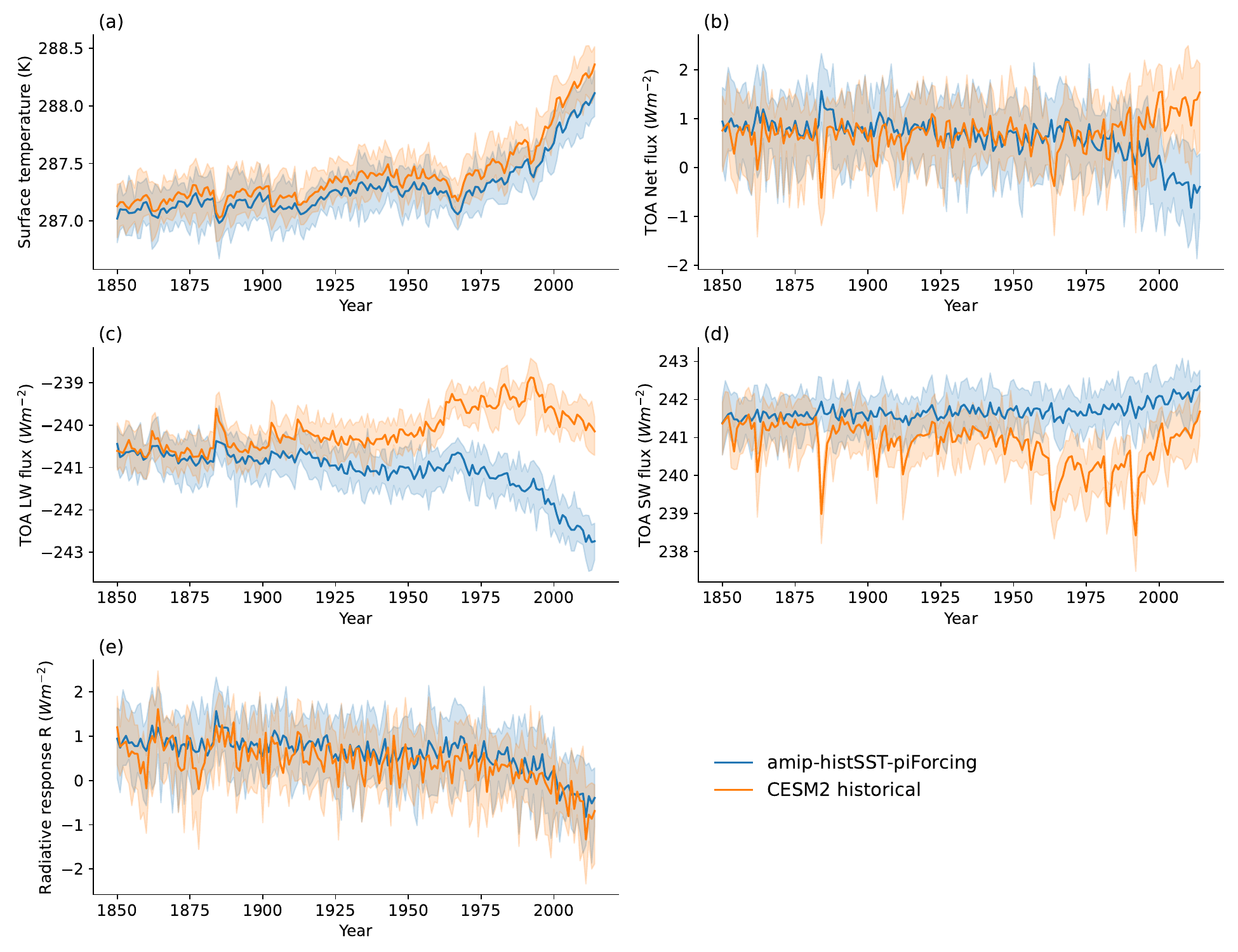}
	\caption{Results from \textit{amip-piForcing-histSST} simulations (blue) compared to the CESM2 \textit{historical} simulations (orange) with pairwise identical SSTs (i.e., \textit{amip-piForcing-histSST} are forced with SSTs from CESM2 historical fully coupled simulations) for (a) global surface air temperature, (b) net TOA imbalance, (c) net TOA LW imbalance, (d) net TOA SW imbalance, and  (e) radiative response $R$. For \textit{amip-piForcing-histSST}, $R$ is identical to $N$, the net TOA imbalance shown in (b). For \textit{historical} runs, $R$ is calculated as $N-F$ (net TOA imbalance as in (b), minus a forcing estimate). $F$ is derived using the RFMIP method (see Appendix~\ref{sec:app:meth_diagnosing_F}). }
     \label{fig:1_basicmetrics}
\end{figure}

We compare global time series of surface air temperature and net, SW, and LW fluxes at the TOA in the fully coupled \textit{historical} simulations and their counterpart \textit{amip-piForcing-histSST} simulations, with pairwise identical SSTs (Fig.~\ref{fig:1_basicmetrics}a--d). The coupled and \textit{amip-piForcing} runs differ only little in their global surface temperature evolution (a), but substantially in the radiative fluxes (b--d). The coupled runs exhibit a slightly higher surface temperature and warming, which has also been found by \cite{andrews_effect_2022} and is well-understood: in the absence of increasing atmospheric CO\textsubscript{2} concentrations, the land temperature, which is not prescribed in \textit{amip-piForcing} runs, is cooler than in the coupled run with forcing. It warms only indirectly due to the warming of the SSTs. \newline

Radiative fluxes exhibit differences between the coupled and \textit{amip-piForcing} setup, which are indicative of the absence of the time-evolving forcing in the \textit{amip-piForcing} runs. In the fully coupled runs, as in the real world \citep{forster_indicators_2025}, the net radiative flux at the TOA becomes increasingly positive in recent decades (Fig.~\ref{fig:1_basicmetrics}b). In contrast, the \textit{amip-piForcing} trend shows a decrease in the net flux, which was also identified by \citet{andrews_using_2014}: this is owed to progressively warmer SSTs, which lead to increased outgoing LW radiation. Since there is no increasing CO\textsubscript{2} concentration in \textit{amip-piForcing}, there is also no additional CO\textsubscript{2} greenhouse effect to compensate the increased outgoing LW radiation. This is confirmed by the increasingly more negative LW flux in the \textit{amip-piForcing} runs (Fig.~\ref{fig:1_basicmetrics}c). In contrast, the LW flux of the coupled run first increases, owing to the trapping of LW radiation from increased CO\textsubscript{2}, and only starts to decrease again once the Planck feedback starts to dominate due to increasingly warmer surface temperatures \citep{myhre_observed_2025}. \newline

\begin{figure}[h] 
	\centering
	\includegraphics[width=\textwidth]{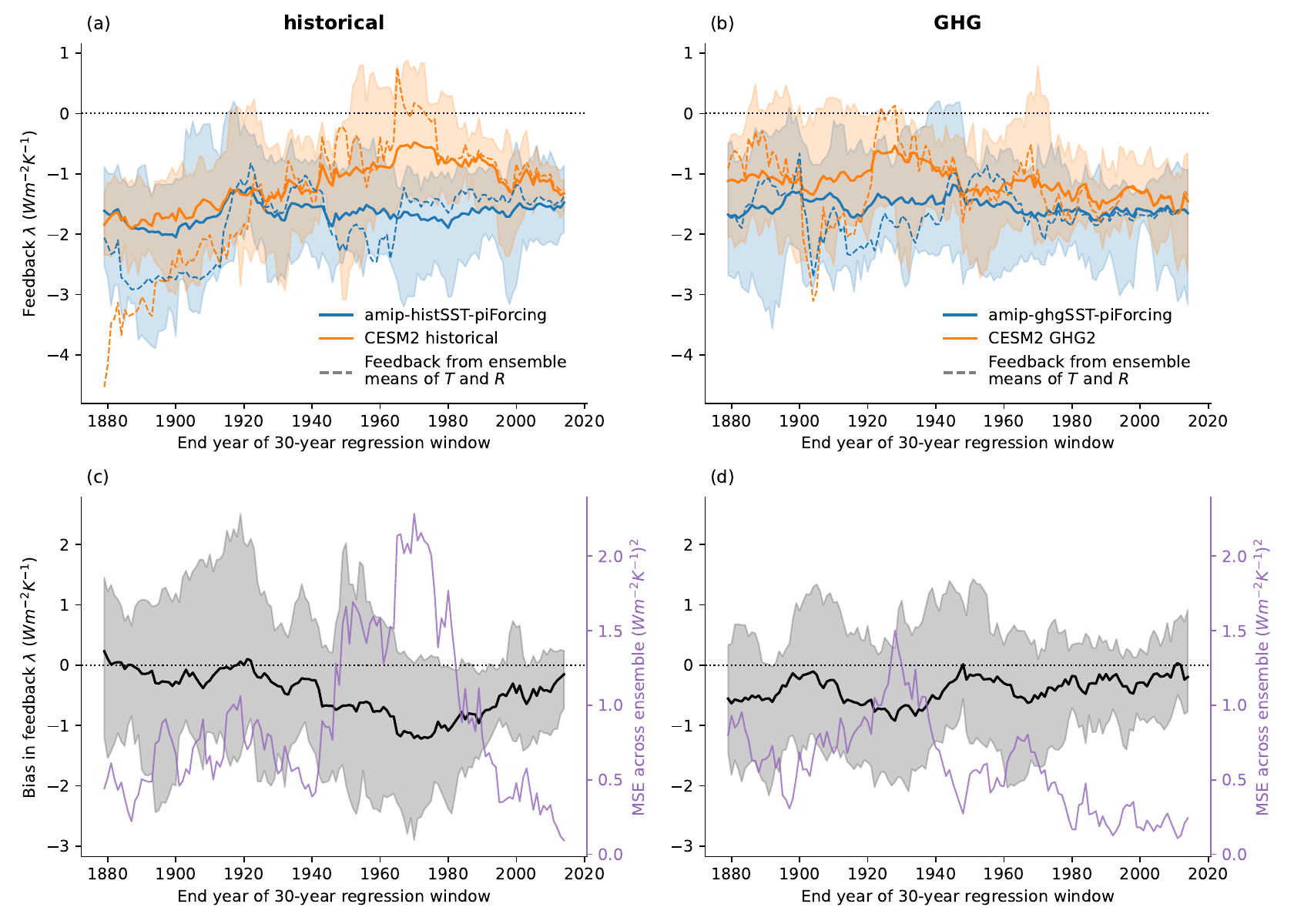}
	\caption{(a) The feedback parameter in \textit{amip-piForcing-histSST} runs (blue) compared to CESM2 \textit{historical} runs with pairwise identical SSTs. Dotted lines show the feedback calculated on ensemble means of $N$ and $T$ (also in (b)). (b) The feedback parameter in \textit{amip-piForcing-ghgSST} (blue) compared to the CESM2 single-forcing GHG2 runs (orange), with pairwise identical SSTs. Panel (c) and (d): black line shows the difference in the feedback between the coupled and atmospheric setup (blue minus orange lines in (a) and (b)) for \textit{historical} and GHG2 SSTs, respectively. The purple line shows the average MSE across the 11 members of the ensemble for the simulations with \textit{historical} SSTs (c), and GHG2 SSTs (d).}
	\label{fig:feedback_amip_vs_cpld}
\end{figure}

In the SW flux, the main difference between \textit{amip-piForcing} and coupled runs hints to the effect of aerosols and volcanoes. The coupled runs exhibit the characteristic dimming and brightening \citep{wild_perspective_2026} along with spikes of lower SW flux following volcanic eruptions. The \textit{amip-piForcing runs}, on the other hand, only exhibit a gradual increase in the incoming SW flux in the last years of the simulation. This increase could be owed to both the decreasing SIC, as well as positive cloud feedbacks which are more prevalent with increasing SSTs in both the coupled and \textit{amip-piForcing} simulations. \newline

The computation of the feedback according to Eq.~\ref{eq:dR/dT} requires the regression of the radiative response $R$, shown in Fig.~\ref{fig:1_basicmetrics}e, onto the surface air temperature in Fig.~\ref{fig:1_basicmetrics}a. At first sight, the differences in $R$ shown in panel e) do not show a large systematic difference between the coupled setup ($R = N - F$) and the \textit{amip-piForcing} setup ($R = N$). The radiative response in the \textit{amip-piForcing} setup lies slightly above the one from the coupled setup, although the difference is small. Combined with the comparatively small difference in the $T$ time series (panel a), one would expect a similar feedback time series (computed as 30-year moving window regressions of $R$ against $T$) between the \textit{coupled} and \textit{amip-piForcing} runs. \newline

The ensemble mean of the feedback time series for \textit{amip-piForcing} and  \textit{historical} coupled runs with pairwise identical SSTs is shown in Fig.~\ref{fig:feedback_amip_vs_cpld}a. The feedback is systematically different in the \textit{amip-piForcing-histSST} and coupled \textit{historical} runs, with the coupled feedback being consistently weaker than the feedback derived from \textit{amip-piForcing} runs (Fig.~\ref{fig:feedback_amip_vs_cpld}a). The average correlation between the coupled and \textit{amip-piForcing} feedback across the ensemble is low (0.23), and the mean squared error (MSE) 0.9~W$^2$m$^{-4}$K$^{-2}$ (Table~\ref{tab:performance_table}). The bias in the feedback estimation is particularly pronounced in the \textit{historical} setup (Fig.~\ref{fig:feedback_amip_vs_cpld}a,c). For the period 1940--2000 the ensemble mean of the coupled feedback shows a pronounced weakening and strengthening, deviating from the \textit{amip-piForcing} feedback by >1~W$^2$m$^{-4}$K$^{-2}$  around 1970. This difference is particularly striking when looking at the similar $T$ and $R$ time series for the \textit{amip-piForcing} and coupled simulations (Fig.~\ref{fig:1_basicmetrics}a,e). The mean squared error between the coupled and \textit{amip-piForcing} feedback across the eleven members also peaks in this period (Fig.~\ref{fig:feedback_amip_vs_cpld}c). The period of largest deviation coincides with the period of strongest aerosol forcing \citep{forster_earths_2021}. This could indicate that the missing aerosol forcing in the \textit{amip-piForcing} causes this discrepancy. \newline

We therefore also compare the difference in \textit{amip-piForcing} and coupled feedback in simulations with historical GHG forcing for the coupled runs, and pairwise identical SSTs in the \textit{amip-piForcing} runs. In the GHG-only setup, the strong systematic deviation in the feedback time series during the 1940--2000 period disappears (Fig.~\ref{fig:feedback_amip_vs_cpld}b,d). However, the consistent negative bias of the \textit{amip-piForcing} feedback remains throughout the entire period (Fig.~\ref{fig:feedback_amip_vs_cpld}b,d). The average Pearson correlation of the coupled and \textit{amip-piForcing-ghgSST} feedback time series increases to 0.39 compared to the setup with historical forcing (Pearson corr.$=$0.23), and the MSE drops from 0.9 to 0.59~W$^2$m$^{-4}$K$^{-2}$ (Table~\ref{tab:performance_table}). The improvement in the GHG-only setup supports the hypothesis that the missing aerosol forcing in \textit{amip-piForcing-histSST} runs is responsible for the large bias around 1940--2000. \newline

\begin{figure}[h] 
	\centering
	\includegraphics[width=\textwidth]{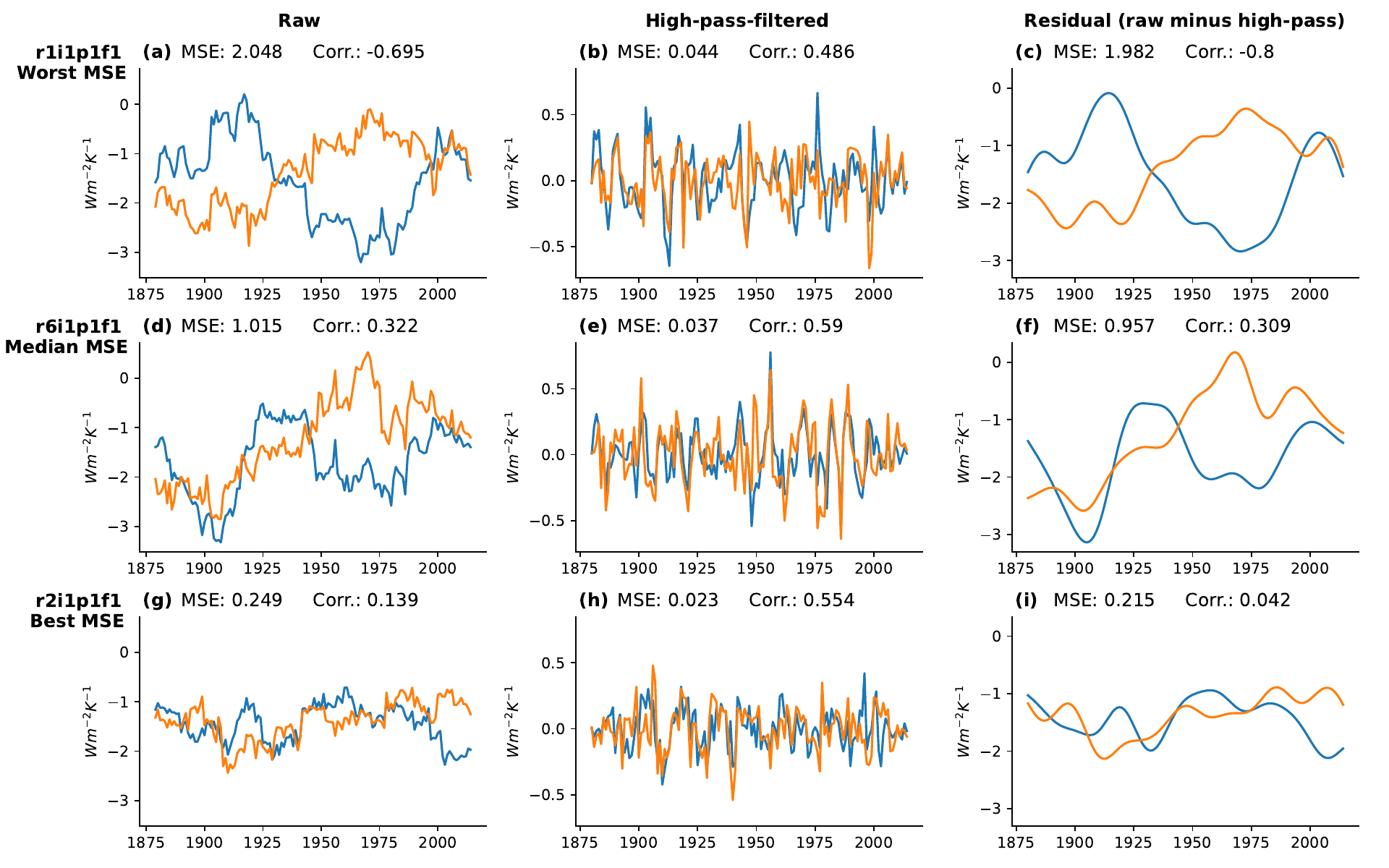}
	\caption{Feedback parameter for three individual members (columns) with pairwise identical SSTs: \textit{amip-piForcing-histSST} simulation (blue) and the corresponding coupled \textit{historical} simulation (orange). Left column (a),(d),(g): the raw feedback parameter time series. Middle column (b), (e), (h): high-pass filter applied to the corresponding raw feedback parameter from the left column. The high-pass filter (Butterworth filter) has a cutoff frequency of 0.05, eliminating variability occurring on timescales >20 years. Right column (c), (f), (i): low-pass filtered feedback (the residual resulting from the subtraction of the high-pass filter from the original feedback parameter time-series). The three members shown are those with the "worst" (top row), "median" (middle row), and "best" (bottom row) MSE between coupled and corresponding atmosphere-only run across the 11-member ensemble.}
	\label{fig:individ_runs}
\end{figure}

We test the sensitivity of our results to the way the feedback parameter is calculated. First, the results remain qualitatively the same if the feedback time series is directly computed on ensemble mean quantities of $T$ and $R$ (dotted lines in Fig.~\ref{fig:feedback_amip_vs_cpld}a,b), rather than averaging the individual feedback time series of individual members. Second, we test the robustness of our results with respect to the forcing estimate $F$ used for computation of the coupled feedback. The results remain largely unchanged if the forcing is either i) corrected for the land surface response using the \citet{hansen_efficacy_2005} method, or ii) computed via the AerChemMIP method \citep{collins_aerchemmip_2017} (Figs.~\ref{fig:app:F-estimates} and \ref{fig:app:F-sensitivity}). This consolidates our findings that the bias in the feedback between coupled and \textit{amip-piForcing}-style runs with identical SSTs and SIC is owed to an inherent difference in the experimental setup.\newline

The results from Fig.~\ref{fig:feedback_amip_vs_cpld} demonstrate that the \textit{amip-piForcing} setup fails to capture the evolution of the feedback in the coupled setting, despite identical SSTs. The particularly large difference in the feedback during the period of strong aerosol forcing suggests that the lack of atmospheric forcing in the \textit{amip-piForcing} simulations might be causing the mis-estimation of the true coupled feedback. \newline

Zooming into individual ensemble members of the \textit{amip-piForcing-histSST} and corresponding \textit{historical} runs reveals large differences between the best- and worst-performing members in terms of mean squared error (MSE) (Fig.~\ref{fig:individ_runs},a,d,g). In the member with the worst MSE (a), the feedback time series happens to be anti-correlated, despite identical SSTs. During the 1960--1980 period the feedback differs by up to 2~Wm$^{-2}$K$^{-1}$ between \textit{amip-piForcing} and \textit{historical} member (Fig.~\ref{fig:individ_runs}a). However, even the member with the lowest MSE (g) shows large discrepancies in the computed feedback time series between \textit{amip-piForcing} and \textit{historical}: for example, the negative trend in the \textit{amip-piForcing} feedback from 1975 onwards, which is similar to the pronounced negative trend found in many \textit{amip-piForcing} experiments using observed SSTs \citep[e.g.,][]{andrews_dependence_2018,andrews_effect_2022,zhou_greater_2021,dong_biased_2021}, is completely absent in the corresponding \textit{historical} fully coupled run, despite identical SSTs and SIC. This implies that the conclusion made in the cited literature above, namely that the negative trend in the feedback is owed to the specific observed SST pattern driving negative feedbacks, is not robust. \newline

We decompose the variability in the feedback time series using a high-pass Butterworth filter \citep{butterworth_theory_1930}, which filters any variability occurring on timescales >20 years (Fig.~\ref{fig:individ_runs}b,e,h). On short timescales, the coupled and \textit{amip-piForcing} feedback time series exhibit much better agreement, with an average Pearson correlation across members of 0.54 and an average MSE of 0.04. Contrarily, the low-pass residual, obtained by subtracting the high-pass filtered time series from the original feedback time series, shows even worse agreement than the original time series. Our results show that the short-term (interannual and decadal) radiative and temperature fluctuations, which are largely governed by SST variability, are mostly reproduced correctly in the \textit{amip-piForcing} runs. This suggests that a slower driver, rather than interannual SST variability, is driving the discrepancy between the coupled and \textit{amip-piForcing} feedbacks. This is unexpected, as long-term variability in $\lambda$ is assumed to come from SST variability, which is, however, the same in both runs. The missing forcing in \textit{amip-piForcing} runs therefore remains a plausible candidate for explaining the discrepancy.  

\begin{table}[htbp]
    \centering
    \scriptsize 
    \caption{MSE and Pearson correlation of feedback time series for paired simulations with identical SSTs but different setup. The table shows how feedback time series of the AMIP-type experiment compare to the feedback time series of the corresponding coupled simulations with identical SSTs and SIC. The MSE and Pearson correlation are averaged across all members and model years.}
    \label{tab:performance_table}
    \renewcommand{\arraystretch}{1.3}
    
    \begin{tabular}{
        >{\raggedright\arraybackslash}p{4.2cm} 
        >{\centering\arraybackslash}p{1.3cm} 
        >{\centering\arraybackslash}p{1.6cm} 
        >{\centering\arraybackslash}p{1.8cm} 
    } 
        \hline 
        AMIP experiment & Target coupled simulation & Ensemble average MSE & Pearson Correlation \\
        \hline 
        \textit{amip-piForcing-histSST}   & \textit{historical} & 0.90 & 0.23 \\ 
        \textit{amip-piForcing-ghgSST}   & \textit{GHG2}& 0.59 & 0.39 \\ 
        \textit{amip-histForcing-histSST}  & \textit{historical} & 0.48 & 0.64  \\ 
        \textit{amip-piForcing-piSST}    & \textit{piControl}& 0.27 & 0.49\\ 
        \hline 
    \end{tabular}
\end{table}

\begin{figure}[h] 
	\centering
	\includegraphics[width=0.95\textwidth]{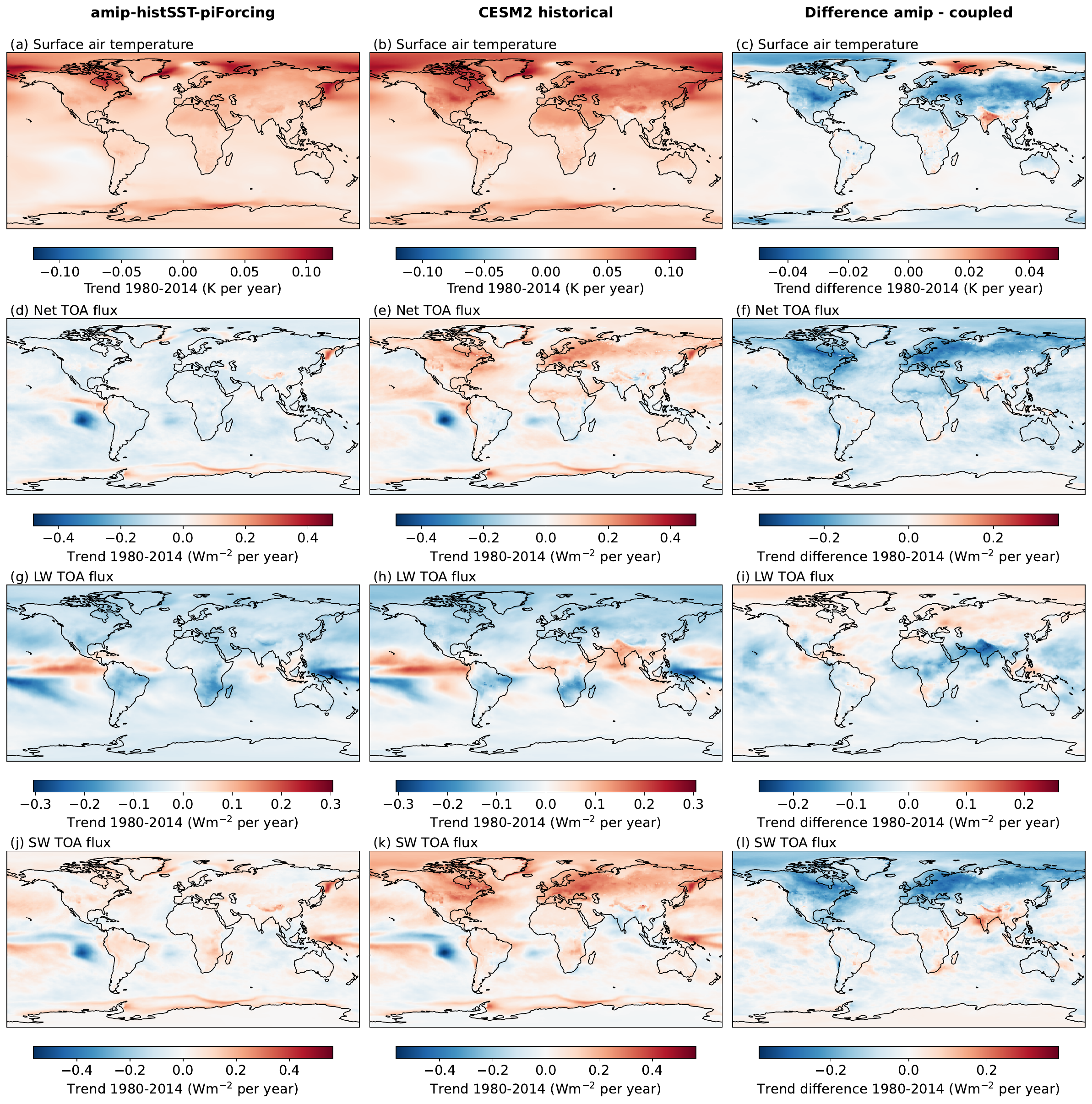}
	\caption{Ensemble mean trends for the period 1980-2014 for the variables surface air temperature (a--c), net TOA flux (d--f), long-wave TOA flux (g--i), and short-wave TOA flux (j--l). The first column shows the trends in the \textit{amip-piForcing-histSST} experiments. The middle column shows the trends in the CESM2 \textit{historical} members whose SSTs are prescribed in the \textit{amip-piForcing-histSST} simulation. The third column shows the difference in the trends computed as \textit{amip-piForcing-histSST} minus \textit{historical}.}
	\label{fig:spatial_trends}
\end{figure}

To gain insight into the spatial structure of differences between \textit{amip-piForcing-histSST} and the coupled \textit{historical} runs we compute temporal trends of surface air temperature ($T$), net TOA flux ($N$), LW TOA flux, and SW TOA flux for the period 1981--2014, for each grid cell (Fig.~\ref{fig:spatial_trends}). Spatial temperature trends show that \textit{amip-piForcing} simulations warm less than the coupled run during this period, particularly over Northern Hemisphere land masses (Fig.~\ref{fig:spatial_trends}a--c). An exception is South-East Asia and the Indian peninsula in particular. This pattern again hints at the effects of aerosol forcing in the coupled simulation, which are absent in the \textit{amip-piForcing} simulations. Aerosol forcing decreases in the 1981--2014 period in most of the Northern Hemisphere apart from South-East Asia, leading to an increased warming signal \citep{forster_earths_2021}. The lower warming in \textit{amip-piForcing} overall and over land is also expected due to the lack of GHG forcing in the \textit{amip-piForcing} runs and the fact that land temperatures are not prescribed in \textit{amip-piForcing}. \newline

The 1981--2014 trends in net TOA radiation differ in sign almost everywhere on the globe (Fig.~\ref{fig:spatial_trends}d--f), as expected from the diverging time series in Fig.~\ref{fig:1_basicmetrics}b. Again, Northern Hemisphere land masses stand out as areas where the \textit{amip-piForcing} trends underestimate the \textit{historical} trends most (Fig.~\ref{fig:spatial_trends}f). The decreasing aerosol forcing in \textit{historical} runs---which is absent from the \textit{amip-piForcing} runs---could explain the stronger net radiation trends in the Northern Hemisphere during this period. This hypothesis is further supported by the SW trends (Fig.~\ref{fig:spatial_trends}j--l), which dominate the net trends: much stronger positive SW TOA imbalance trends are found in Europe and North America, whereas South-East Asia---which does not experience aerosol reductions in the 1981--2014 period but an increase \citep{yang_global_2024}---even shows slightly negative SW EEI trends for the same period. The LW imbalance (Fig.~\ref{fig:spatial_trends}g--i) also aligns with this interpretation: while the sign of trends is negative almost everywhere on the globe in both simulation setups, the \textit{amip-piForcing} simulations show a less negative LW trend over the Northern Hemisphere continents, with the exception of South-East Asia. Again, if decreasing aerosols lead to a stronger warming trend over the NH land masses in the coupled runs, this could explain the stronger negative LW trends there compared to the \textit{amip-piForcing} trends. \newline

Our results remain qualitatively the same if we only consider trends over the boreal summer months or for clear-sky fluxes (Figs.~\ref{fig:app:spatial_clearsky},~\ref{fig:app:jja_spatial}).  If snow albedo effects were the main contributor to the differences, we would not expect to see this effect in the boreal summer months.  If the differences were mainly owed to cloud effects, we should not expect to see the same signature also in clear-sky fluxes. This consolidates the hypothesis that the missing $F$ in \textit{amip-piForcing} runs explains most of the spatial differences in Fig.~\ref{fig:spatial_trends}.\newline

\begin{figure}[h] 
	\centering
	\includegraphics[width=0.8\textwidth]{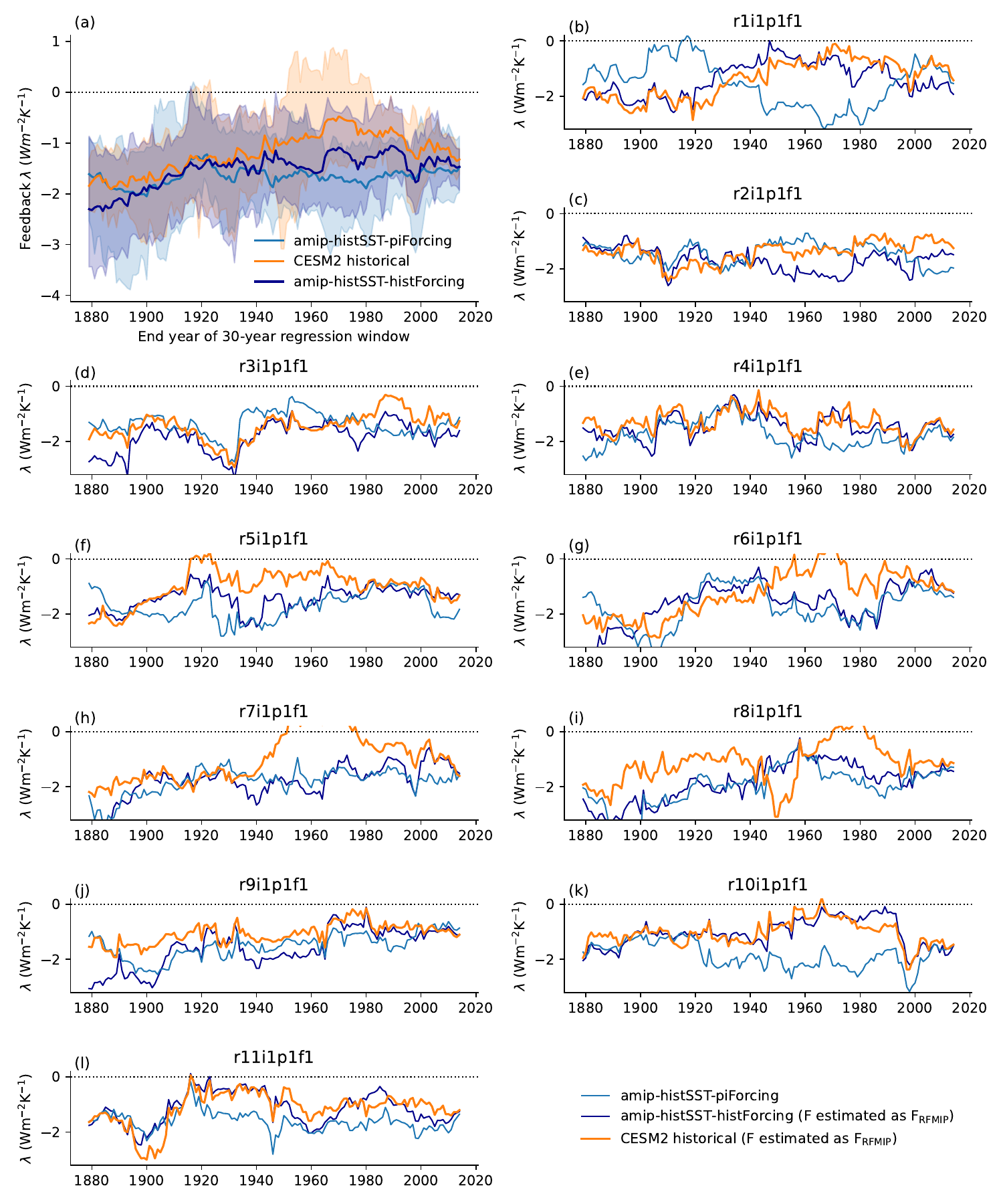}
	\caption{(a): like Fig.~\ref{fig:feedback_amip_vs_cpld}a, but with the ensemble mean and spread from \textit{amip-histForcing-histSST} included in dark blue. (b-l): feedback parameter $\lambda$ for individual members, all members in one panel have identical SSTs. Light blue lines show the \textit{amip-piForcing-histSST} feedback, as in Fig.~\ref{fig:individ_runs}a,d,g. Orange line shows the feedback in the corresponding coupled historical run, and the dark blue line shows the feedback estimated from \textit{amip-histForcing-histSST} runs, recovered via Eq.\ref{eq:general_feedback} with the forcing $F$ estimated as $F_{\mathrm{RFMIP}}$.}
	\label{fig:feedbacks_amipHistForcing}
\end{figure}

We now test the hypothesis that the absence of time-varying atmospheric forcing in \textit{amip-piForcing} simulations can explain the bias in the feedback calculation. The feedback calculated from \textit{amip-histForcing-histSST} runs---which are identical to the \textit{amip-piForcing-histSST} runs but now have time-varying prescribed \textit{historical} radiative forcing---is shown in Fig.~\ref{fig:feedbacks_amipHistForcing} (dark blue lines). On average, the Pearson correlation between the feedback in \textit{amip-histForcing} and coupled \textit{historical} runs shows a large improvement to 0.64, compared to 0.23 for the \textit{amip-piForcing} runs (Table~\ref{tab:performance_table}), while the MSE is reduced from 0.90~Wm$^{-4}$K$^{-2}$ to 0.48~Wm$^{-4}$K$^{-2}$. However, there remains a negative bias of the \textit{amip-histForcing-histSST} ensemble mean compared to the coupled ensemble mean during the 1940--2000 period, albeit reduced compared to the \textit{piForcing} setup (Fig.~\ref{fig:feedbacks_amipHistForcing}a). For some members (e.g., Fig.~\ref{fig:feedbacks_amipHistForcing}b,~d,~e,~k), the \textit{amip-histForcing} feedback indeed tracks the coupled feedback much better (orange lines), however this is not the case for others (e.g., Fig.~\ref{fig:feedbacks_amipHistForcing}c,~g,~h,~i). \newline 

The strong improvement in the average agreement between coupled and \textit{amip-histForcing} feedback (compared to coupled and \textit{amip-piForcing} feedback) suggests that the assumption $F=0$ made in \textit{amip-piForcing} runs leads to a substantial bias in the feedback estimation (Table~\ref{tab:performance_table}). In particular, the spatial analysis in Fig.~\ref{fig:spatial_trends}, the characteristic increase and decrease in the mean bias of the feedback during 1940--2000 (Fig.~\ref{fig:feedback_amip_vs_cpld}a,c), and the reduced bias in the GHG-only setup (Fig.~\ref{fig:feedback_amip_vs_cpld}b,d) strongly hint to aerosols being particularly relevant in causing the discrepancy. Aerosols are a) fairly short-lived and b) are generated over land: this leads to the fact that only a small part of the aerosol forcing is reflected in the SSTs. GHGs, in contrast,  are well-mixed, long-lived, and have been increasing over the entire historical period, so their radiative forcing has had time to be reflected in the SSTs. Because only SSTs---but not land temperatures---are prescribed in the \textit{amip-piForcing} setup, forcing effects which are not reflected in the coupled runs' SSTs will not end up in the \textit{amip-piForcing} run. \newline

Nonetheless, it is remarkable to see that even with $F$ prescribed identically as in the coupled \textit{historical} runs, there remain substantial differences in the feedback time series (e.g., Fig.~\ref{fig:feedbacks_amipHistForcing}a,~b,~f,~g). In these runs there are periods where the feedback time series diverge substantially despite identical prescribed SSTs and atmospheric forcing. This suggests that something other than SSTs or forcing is driving these differences. In particular, it suggests that SST and SIC alone are unable to fully constrain the feedback response in simulations with identical SSTs and SIC. This finding implies that inferring Earth's feedback from \textit{amip-piForcing} simulations with prescribed observed SSTs, SIC and atmospheric forcing may still not provide an unbiased estimate of Earth's observed feedback evolution.\newline

The remaining differences between \textit{amip-histForcing-histSST} and coupled \textit{historical} simulations could arise for two reasons. First, it is possible that there are sources of variability in the feedback time series, which are unrelated to the SST or SIC pattern. Such sources could be, for example, physical variability originating over land or the atmosphere and unconstrained by SSTs, or statistically-induced variability due to the computation method of $\lambda$ via moving-window regressions. In the following section, we turn to an idealized pre-industrial control setting without potentially confounding radiative forcing in order to investigate such sources of variability in the feedback estimates. Second, it is possible that the \textit{amip}-type simulation does not capture the response of the coupled system correctly. This has been demonstrated in a simple stochastic model and verified in early generation GCMs already by \citet{barsugli_basic_1998} and the follow-up work of \citet{bretherton_interpretation_2000}. In these works, the authors show theoretically and with evidence from GCMs that the severed coupling between atmosphere and ocean leads to unphysically enhanced atmospheric variance and surface heat fluxes. Other works have expanded on these findings and shown that ocean-atmosphere coupling matters for accurate representation of ENSO-teleconnections, for example \citep{alexander_atmospheric_2002}. Whether the bias between \textit{amip}-type and coupled simulations seen in Fig.~\ref{fig:feedbacks_amipHistForcing}a is a result of the missing atmosphere-ocean coupling in \textit{amip}, or is merely an effect of the small sample size, remains unclear and is beyond the scope of this study. 

\subsection{Sources of variability in the feedback parameter}\label{sec:results_Variability}
\begin{figure}[h] 
	\centering
	\includegraphics[width=\textwidth]{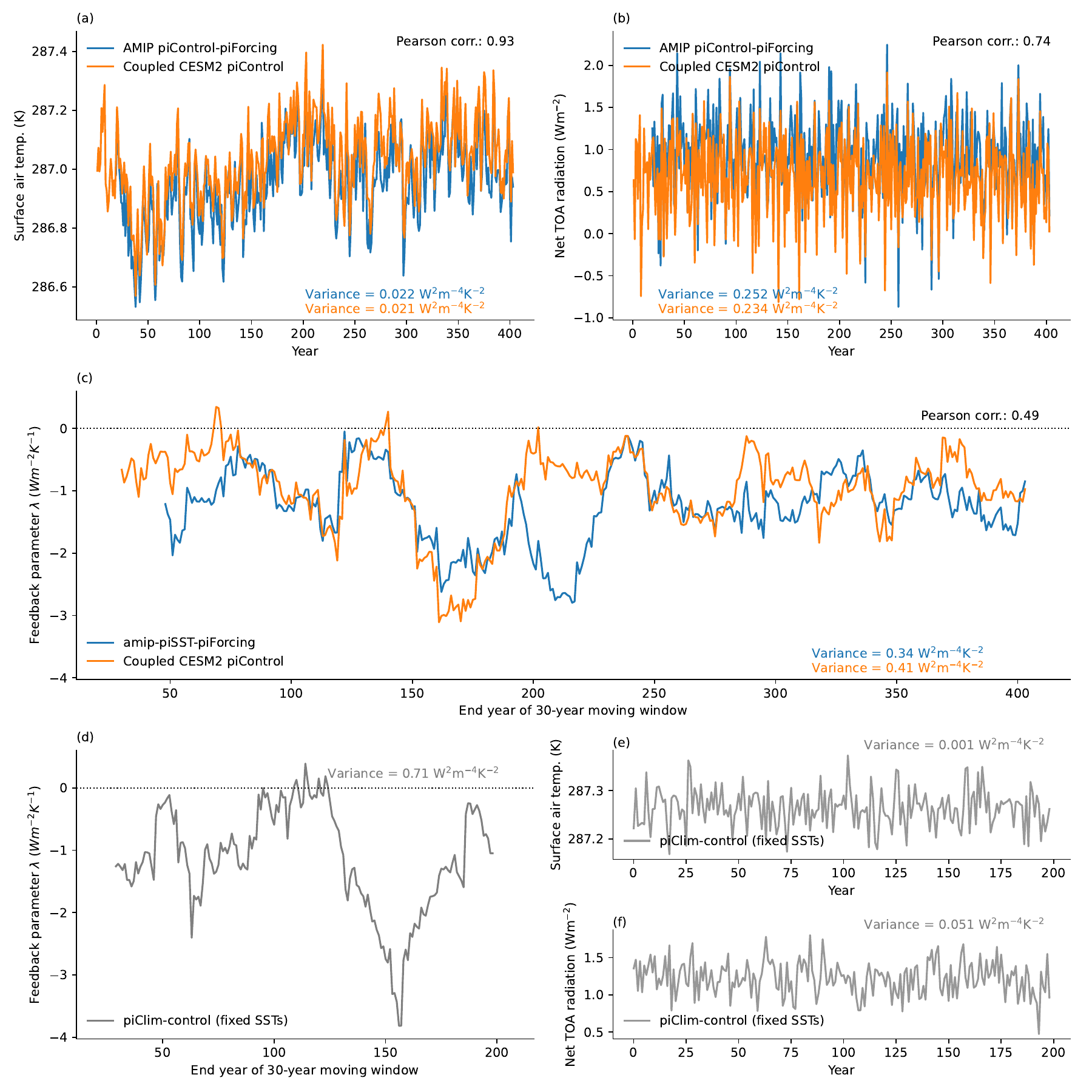}
	\caption{Feedback calculated in the \textit{amip-piForcing-piSST} run (blue) compared to the coupled \textit{piControl} run with identical SSTs (orange). }
	\label{fig:piControl}
\end{figure}

We now turn to idealized \textit{piControl} simulations in order to investigate sources of variability in the feedback parameter time series which could potentially explain the differences between coupled and \textit{amip}-type simulations despite identical SST, SIC and forcing. Fig.~\ref{fig:piControl}a-b show the surface air temperature ($T$) and net TOA radiation ($N$) in the coupled CESM2 \textit{piControl} simulation (orange) and its \textit{amip-piForcing-piSST} counterpart (blue), which has been forced with the SSTs and SIC of the coupled run. The time series of $T$ and $N$ are strongly correlated between \textit{amip-piForcing} and coupled run (Pearson corr.: 0.93 for $T$ and 0.74 for $N$). When the feedback is computed from the time series of $N$ and $T$, however (Fig.~\ref{fig:piControl}c), the correlation drops to 0.49. Although the Pearson correlation of 0.49 between the two feedback time series is stronger than the correlation found between the \textit{amip-piForcing-histSST} and corresponding \textit{historical} runs (0.23), there nonetheless remain substantial differences in the estimated feedback. Notably, there are periods---such as years 30--80, or years 200--240 of the simulation---during which the feedback time series diverge substantially. How can this be the case, when the AMIP version of $T$ and $N$ seems to follow the coupled $T$ and $N$ so closely, with no apparent bias by visual inspection? \newline

Our results suggest that minor variations in $N$ and $T$ between the coupled and \textit{amip}-style simulations, which are not obvious from visual inspection (Fig.~\ref{fig:piControl}a,~b), become amplified through the feedback calculation and result in feedback discrepancies over large periods. This must be the case, because if there were substantial differences in the $N$ and $T$ between the two setups, we should have been able to see them in Fig.~\ref{fig:piControl}a,~b. In addition, AMIP simulations have generally been routinely tested in terms of their capability to reproduce global variables like net TOA radiation and temperature for a long time \citep[e.g.,][]{gates_overview_1999}. Therefore, we hypothesize that what we are seeing is a statistical artifact of the moving window regressions, where small deviations in the $N$ and $T$ time series---for instance due to unconstrained variability over land regions and in the atmosphere---might lead to large periods in the moving-window feedback time series, where the \textit{amip-piForcing} and coupled simulations diverge. \newline

\subsubsection{The Yule-Slutzky effect: how random noise generates multi-decadal variability}\label{sec:res:yule-slutzky}
To test this hypothesis, we turn to the \textit{piClim-control} simulation, where SSTs and SIC are held climatologically fixed and vary only with the seasonal cycle (Fig.~\ref{fig:piControl}d--f). First, the output of $T$ and $N$ (Fig.~\ref{fig:piControl}e,~f) reveals that $N$ and $T$ have a much smaller variance than in the coupled \textit{piControl} simulation, and in particular, do not seem to exhibit long-term variability. The joint autocorrelation function of $T$ and $N$ is not significant at any lag for any variable, exactly as expected from a prescribed climatological SST-setup, where memory from the ocean is erased. Also, the distributions of $N$ and $T$ (and their joint distribution) are both consistent with white noise, which we tested via uni-variate \citep{dagostino_tests_1973} and multi-variate \citep{henze_class_1990} normality tests (see Sec.~\ref{sec:meth:montecarlo}). \newline 

\begin{figure}[h] 
	\centering
	\includegraphics[width=\textwidth]{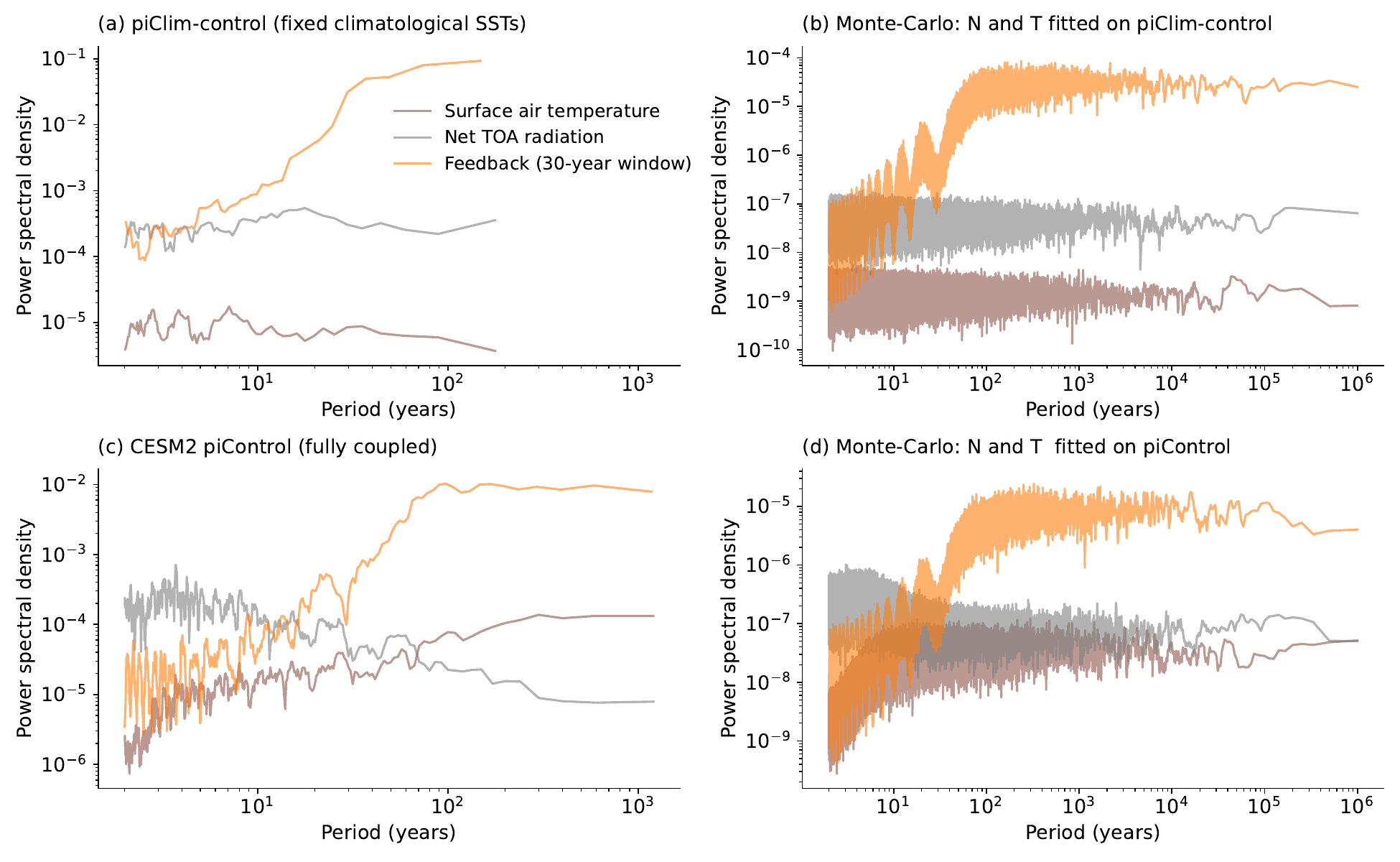}
	\caption{Spectral densities of the variables $T$ (surface air temperature), $N$ (net TOA imbalance) and the feedback $\lambda$, computed via 30-year moving window regressions of the $N$ and $T$ timeseries (Eq.~\ref{eq:amip_pf_movwin}), (a): for the 200-year long \textit{piClim-control} simulation, (b): for synthetic timeseries generated via a Monte-Carlo simulation. $N$ and $T$ were fitted on data from \textit{piClim-control} as jointly normal with no autocorrelation (see Sec.~\ref{sec:meth:montecarlo}); (c): for the 1200-year long coupled \textit{piControl} simulation, (d): for synthetic time series generated via a Monte-Carlo simulation. $N$ and $T$ were fitted on data from \textit{piControl} as a VAR(5) process (see Sec.~\ref{sec:meth:montecarlo}).}
	\label{fig:montecarlo}
\end{figure}

However, the feedback time series computed from $T$ and $N$ in \textit{piClim-control} still exhibits substantial interannual variability (Fig.~\ref{fig:piControl}d), with periods of lower or higher feedback, similar to the \textit{piControl} feedback time series (Fig.~\ref{fig:piControl}c). We illustrate this further by looking at the spectral densities of $N$, $T$, and the feedback in \textit{piClim-control} (Fig.~\ref{fig:montecarlo}a): the spectral density of $N$ and $T$ is rather flat, as expected for white noise time series, but the power spectral density of the feedback time series increases substantially towards lower frequencies. This is noteworthy, as the feedback has been computed from two time series which both exhibit no significant autocorrelation and are consistent with white noise. \newline

We want to point out that observing variability in the \textit{piClim-control} feedback of a magnitude similar to the \textit{piControl} feedback is not surprising per se: since the feedback is computed as a ratio between $N$ and $T$, it is independent of the absolute magnitude of variability in $N$ and $T$. What matters for the feedback is only the variability of $T$ and $N$ relative to one another. In other words, even small variations in $T$ and $N$ could result in the same feedback variance, if the relative variability of $T$ and $N$ is the same. In fact, we would even expect a higher variance in the feedback in an unforced (such as \textit{piControl}) or even fixed-SST setup (such as \textit{piClim-control}): the absence of forcing implies that over the 30 years used for the computation of $\lambda$ there is a weak signal in the denominator ($\Delta T$), which, if $\Delta T$ comes close to zero, inflates the variance of the feedback timeseries. This is indeed reflected in the high variance of the \textit{piClim-control} feedback (Fig.~\ref{fig:piControl}d). It also provides a reason why the correlation between coupled and \textit{amip} feedback is worse for the \textit{piControl}-case than for the \textit{amip-histForcing-histSST} case, as the latter has a historical prescribed forcing causing a potentially stronger signal in $\Delta T$. What is striking, however, is that the variability spectrum of the \textit{piClim-control} feedback exhibits long-term variability with substantial memory (as seen from the increasing power spectral density towards lower frequencies in Fig.~\ref{fig:montecarlo}a), or, as seen in Fig.~\ref{fig:piControl}d from pure visual inspection: sustained periods of weaker or stronger feedback despite there being no memory in the generating time series $N$ and $T$.\newline

What we observe in Figs.~\ref{fig:piControl}d--f and \ref{fig:montecarlo}a---namely, two white noise time series generating low-frequency variability after applying a moving-window regression---is an example of what is known as the Yule-Slutzky effect \citep{slutzky_summation_1937, yule_why_1926, gershunov_low-frequency_2001,kelly_change_2014}: when a moving window is applied over a random noise process, it generates long-term variability which does not stem from actual low-frequency variability of the underlying data. The effect of applying moving window averages on noisy climate data has been documented and studied before \citep[e.g.,][]{gershunov_low-frequency_2001,kelly_change_2014,li_enhanced_2024}, yet moving averages nonetheless remain a common and ubiquitous (but often problematic) practice in climate science. \newline 

We perform a Monte-Carlo simulation to show how the Yule-Slutzky effect operates in \textit{piClim-control} and to confirm that the effect is not a result of the relatively short \textit{piClim-control} simulation. For this, we simulate a $1\,000\,000$-year long $T$ and $N$ time series from a joint normal distribution fitted to the \textit{piClim-control} $N$ and $T$ output (see Sec.~\ref{sec:meth:montecarlo}). We then compute the resulting feedback time series as $\frac{\partial N}{\partial T}$ in 30-year moving window regressions. The resulting spectra of $N$, $T$, and the feedback are shown in Fig.~\ref{fig:montecarlo}b. These confirm the results from Fig.~\ref{fig:montecarlo}a: despite the flat white-noise spectra of $N$ and $T$, there is substantial variability of the feedback parameter time series increasing towards longer timescales. The feedback spectrum flattens out at a decorrelation timescale of roughly 60--100 years. This means that there are substantial variations in the \textit{piClim-control} feedback time series up to 60--100 year timescales. Importantly, this multi-decadal variability is purely stochastically generated (in the case of Figure~\ref{fig:montecarlo}b), or results from short-term white-noise-like internal variability in $N$ and $T$ (in the case of Figure~\ref{fig:montecarlo}a). It therefore does not originate from any physical long-term variability or memory in $N$ or $T$. \newline

The idealized \textit{piClim-control} setup helped to identify a purely statistical source of variability in the feedback time series. However, the real world and the fully coupled \textit{piControl} simulation are different from the idealized \textit{piClim-control} simulations, because the freely-evolving ocean-atmosphere system in \textit{piControl} harbors substantial memory and cannot be modeled as a white noise process. This can be seen in the power spectral densities of $N$, $T$, and the feedback from the 1200-year long coupled \textit{piControl} simulation (Fig.~\ref{fig:montecarlo}c). Here, the spectrum of $T$ shows increasing variance up to centennial timescales. This is the expected effect of slow ocean variability leading to strong memory in the system and is typical for time series of climatic variables \citep{hasselmann_stochastic_1976,frankignoul_stochastic_1977,barsugli_basic_1998,roe_feedbacks_2009}. The time series of $N$, on the other hand, shows a power spectrum which peaks at a frequency of $\sim 2-5$ years, which is expected given this roughly corresponds to the ENSO timescale. The power spectral density then decreases with period length, indicating a process with maximal variability on the shortest timescales and highly stable at long timescales. This spectrum of $N$ is also to be expected, and follows directly from a linear coupled system with stochastic forcing and a restoring, negative feedback: at long timescales, the restoring feedback ensures that the system restores equilibrium ($N=0$), meaning that long periods with higher or lower $N$ do not occur internally and hence, on long timescales, $N$ is constant \citep{barsugli_basic_1998,bretherton_interpretation_2000}. \newline

The feedback spectrum of the \textit{piControl} simulation (Fig.~\ref{fig:montecarlo}c) shows a similar increase at long frequencies like the one of the \textit{piClim-control} simulation (Fig.~\ref{fig:montecarlo}a). We again verify this effect via a Monte-Carlo simulation. This time, we model the joint evolution of the coupled \textit{piControl} \textit{N} and \textit{T} time series via a vector autoregressive (VAR) process of order 5, and again simulate one million years of $N$ and $T$ from it (Sec.~\ref{sec:meth:montecarlo}). The spectra of the simulated $N$ and $T$, as well as the spectrum of the resulting feedback, are shown in Fig.~\ref{fig:montecarlo}d. The results reproduce the spectrum obtained from the raw \textit{piControl} simulation. The variance of the feedback calculated from the 30-year moving window regressions peaks at periods of 60--100 years and longer. This means that also in the coupled simulation, any trends in the feedback detected on timescales shorter than $\sim 100$ years could be owed to statistical artifacts, making their attribution to changing physical mechanisms (such as the evolving SST pattern) extremely difficult. \newline 

That being said, the Monte-Carlo simulation using the VAR(5) process fitted on coupled \textit{piControl} data does not say anything about the origin of the statistically found (auto-)correlations in $N$ and $T$. It only demonstrates that the Yule-Slutzky effect could be generating substantial variability also in the non-white-noise setting of the coupled simulation, accounting for some of the feedback variance therein. The results in Fig.~\ref{fig:montecarlo}c--d do not, however, exclude that the specific autocorrelation structure of $N$ and $T$ found via the VAR(5) process arises from specific, persistent or evolving, patterns of SSTs, which then also drive $N$ and $T$ fluctuations. In other words, our Monte-Carlo simulation is unable to attribute the amount of variance found in the feedback to a specific cause: pure statistical Yule-Slutzky effect, versus a physically meaningful evolution of SST patterns, for instance.\newline

\subsubsection{Is there an unforced SST pattern effect in \textit{piControl}?}\label{sec:res:bootstrap_piControl}

\begin{figure}[h] 
	\centering
	\includegraphics[width=0.85\textwidth, height=0.7\textheight, keepaspectratio]{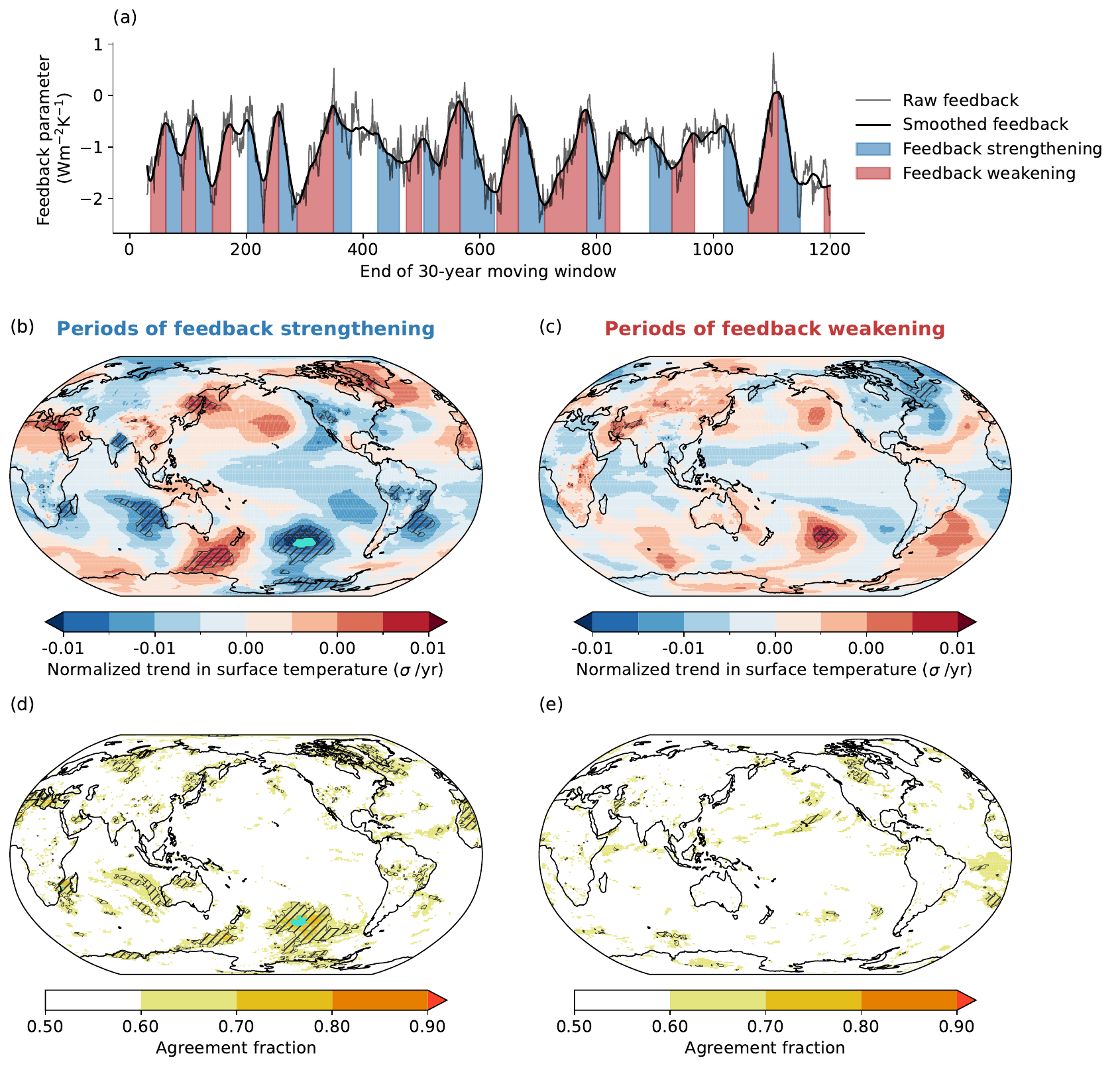}
	\caption{Surface air temperature trend composites in the fully coupled \textit{piControl} run during periods of feedback strengthening and weakening. (a): the raw feedback time series (thin grey line), together with a smoothed feedback time series (thick black line), smoothed using a Savitzky-Golay filter with window length 81 and polynomial order 4. Blue and red shaded regions indicate consecutive periods of length $>10$~years, during which the derivative of the smoothed feedback time series is negative or positive, respectively. A positive derivative means that the feedback is weakening, a negative derivative, that the feedback is strengthening. Panels (b) and (c) show the surface air temperature trend composites computed over the blue (b) and red (c) shaded regions in (a). Panels (d) and (e) show the fraction of feedback strengthening (d) or feedback weakening (e) periods which agree on the sign of the surface air temperature trend in a given location. Black hatching indicates regions where the composite trend (b--c) or agreement (d--e) is significant at the 95\% confidence level. Turquoise areas in (b)--(e) indicate regions significant at the 95\% confidence level after accounting for multiple testing (see Sec.~\ref{sec:meth:bootstrap}).}
	\label{fig:composites}
\end{figure}
In order to investigate whether some of the variations in the coupled \textit{piControl} feedback arise from evolving SST patterns, we investigate periods of strengthening and weakening of the feedback time series in the 4000-year long, \textit{piControl-long} simulation (see Sec.~\ref{sec:meth:bootstrap}). If the SST pattern was driving these fluctuations, we would expect that SST patterns share certain similarities during periods of feedback strengthening or weakening. For instance, a strong East-West Pacific SST gradient has been found to induce more negative feedbacks due to direct effects on lapse rate, inversion strength and clouds \citep[e.g.,][]{ceppi_relationship_2017,andrews_dependence_2018,dong_attributing_2019}. We therefore identify periods of at least 20 consecutive years in the \textit{piControl-long} run, during which the feedback is strengthening or weakening (the first 1200 years of \textit{piControl-long} are shown in Fig.~\ref{fig:composites}a). The resulting composite maps of surface temperature trends during periods of feedback strengthening and weakening are displayed in Fig.~\ref{fig:composites}b and \ref{fig:composites}c. \newline

The surface temperature trend composite maps do not exhibit the patterns that would be expected from literature: the tropics and the tropical Pacific in particular do not stand out with strong trends, although they are believed to be a main driver of feedback variations \citep[e.g.,][]{zhou_impact_2016,ceppi_relationship_2017,andrews_dependence_2018,dong_attributing_2019}. In fact, only very few of the trends in the composite maps are significant at the 5\% level (black hatching), and they are not in the tropics. In particular, there is no significant trend in the Pacific SST gradient. This leads to the question, are any of these trends physically meaningful, and how can our results be reconciled with previous literature? \newline

First, we find that when accounting for multiple testing \citep{benjamini_controlling_1995}, none of the computed trends are significant at the 5\% level apart from a small region in the South Pacific during periods of feedback strengthening (turquoise blob in Fig.~\ref{fig:composites}b). This implies that we are unable to identify a characteristic trend pattern in surface air temperature emerging during periods of feedback strengthening or weakening. We confirm this finding by looking at the agreement on the sign of the trends among periods of weakening or strengthening. We find that the vast majority show very weak agreement (Fig.~\ref{fig:composites}d,~e): the agreement fraction rests below 0.6 almost everywhere, for both the feedback weakening and feedback strengthening periods. This means that in roughly half of the red or blue periods in Fig.~\ref{fig:composites}a, the trend in a region is positive, and in the other half of the periods, it is negative. In other words, the individual periods making up the composite do not agree on the sign of the trend, almost anywhere on the globe. In fact, also for the agreement fraction, we find that almost none of the hatched regions are indeed significant if the multiple testing problem is accounted for: the areas which show an agreement on the sign of the SST trend which is significantly different from random noise at the 95\% level amount to only 6\% of the total number of grid cells for periods of feedback strengthening, for instance---close to the amount expected from a multiple testing error at the 95\% significance level.\newline

This finding is particularly important when considering the discrepancy between observed and modeled SST patterns in the Pacific \citep[e.g.,][]{olonscheck_broad_2020,wills_systematic_2022}. The prevailing consensus in literature is that the strengthening SST gradient in observations is the cause of the stabilizing feedback in recent decades found via \textit{amip-piForcing} simulations \citep{zhou_analyzing_2017,andrews_accounting_2018,dong_biased_2021,andrews_effect_2022}. However, our results suggest that---at least in the unforced \textit{piControl} setting in this climate model---trends in a specific region of the globe cannot be uniquely mapped to a specific feedback evolution. This could either be because (i) the variance in the feedback computed from 30-year moving window regressions is to a large degree statistical noise from the Yule-Slutzky effect, or (ii) it could be because the co-evolution of the SST field over the entire globe is never pronounced enough to lead to detectable changes in the feedback, or (iii) in an unforced setting it is hard to detect the effect of the SST pattern due to less-pronounced signals in $N$ and $T$. In other words, even if a certain region exhibits feedbacks which are known to be stabilizing---such as the recent SST evolution in the tropical Pacific---, this effect could be compensated by opposite effects in other areas of the globe. This would make sense from a statistical point of view, given the multidimensionality of the spatial field and the multitude of feedbacks acting everywhere simultaneously. Therefore, unless the SST pattern in observations is really exceptionally different from what would be expected `naturally', and acts in a direction which affects the feedback positively or negatively in several regions at the same time, our results suggest that there is not enough evidence to claim an effect of the observed SST pattern on the evolution of the observed feedback time series. \newline 

\section{Discussion}
Some of our results may at first appear contradictory to previous literature. Numerous studies show that there are key regions on the globe, which drive important cloud, lapse-rate and water vapour feedbacks, particularly the tropical and subtropical Pacific Ocean \citep{zhou_impact_2016,ceppi_relationship_2017,andrews_dependence_2018,andrews_accounting_2018,dong_attributing_2019}. In addition, it has been shown that in abrupt 4xCO\textsubscript{2}-increase experiments, the shift of the average pattern of warming leads to a change in the feedback towards less stabilizing values \citep[e.g.,][]{senior_time-dependence_2000,andrews_dependence_2015,rugenstein_dependence_2016,dong_biased_2021}. Why do we then not see any of these patterns emerging in the variability of the feedback in the fully coupled \textit{piControl} simulation? We attempt a systematic reconciliation, which is mainly related to the signal-to-noise ratio of the experimental setups.

\subsection{Transient feedback estimates from \textit{amip-piForcing} runs are likely not robust}
Our results suggest that the findings of numerous studies using 30-year moving window regressions to compute the time-evolution of the global feedback based on \textit{amip-piForcing} runs \citep[e.g.,][]{gregory_variation_2016,zhou_impact_2016,andrews_accounting_2018,dong_biased_2021,andrews_effect_2022} may not be robust, for two reasons: first, the lack of prescribed \textit{historical} atmospheric forcing may induce a large bias and poor correlation with respect to the true feedback evolution. Second, the recent period of the past hundred years over which the feedback evolution is evaluated is too short to separate any changes in the feedback from the timescale of statistical noise induced by the Yule-Slutzky effect. Our results suggest that the stabilization of the historical feedback found in these experiments cannot be robustly attributed to changing SST patterns, and that this stabilization may in fact be purely statistical noise. This problem is further aggravated by the fact that \textit{amip-piForcing} runs with observed SSTs and SIC prescribe model-foreign SST boundary conditions to the atmospheric model. This in itself might introduce a bias of unknown magnitude whose quantification is beyond the scope of this paper. The fact that even in our controlled setup---where model-internal SSTs and SIC are prescribed to the same atmospheric model---the feedback of the \textit{amip} setup does not match that of the coupled setup, raises serious concerns about any conclusions on the feedback inferred from \textit{amip-piForcing} with observed SSTs, which (due to structural model biases) will differ from those "preferred" by (i.e., internally consistent with) the atmospheric model. 

\subsection{The signal-to-noise problem in quantifying a  transient, time-evolving pattern effect}
Our findings do \textit{not} challenge fundamental theoretical arguments and model results on how specific SST patterns in specific locations affect cloud, lapse-rate and water vapour feedbacks \citep[e.g.,][]{senior_time-dependence_2000,andrews_dependence_2015,rugenstein_dependence_2016,andrews_dependence_2018}. Rather, our results point to a signal-to-noise problem: In studies where a strong, persistent, idealized perturbation is imposed over a sustained period---such as abrupt-CO\textsubscript{2}-quadrupling experiments \citep[e.g.,][]{andrews_dependence_2015}, AMIP-4K-type experiments \citep[e.g.,][]{andrews_dependence_2018}, or idealized prescribed heat-flux experiments \citep[e.g.,][]{rugenstein_dependence_2016}---there is a much higher signal-to-noise ratio. In these cases, one considers either a long-term mean response (such as the CO\textsubscript{2}-quadrupling), or a response to a fixed SST pattern which is not allowed to (co)vary in time. In both cases, there is much less noise generated, because there is no need to compute moving-window regressions in order to estimate a transient feedback response. The feedback can be directly estimated via the Gregory-method. These sustained SST patterns then may very well impact the feedback, as has been shown. The fact that we do not find said relationships in the transient \textit{piControl} setup is likely due to the covariance of the entire SST field, where compensating effects may occur. In other words, the signal-to-noise ratio in an unforced \textit{piControl} setup is too low to detect the expected physical mechanisms (or, the statistical noise in the transient moving-window calculation is too big). \newline

Also related to the signal-to-noise problem are studies based on Green's function: here, individual patches are systematically subjected to SST perturbations and their effect is therefore spatially isolated, assuming linearity of the total response with respect to the individual geographic responses \citep{zhou_analyzing_2017,dong_attributing_2019}. Green's functions can very well be used to identify the effects of individual warming patterns on the global feedback. However, unless the entire spatially-varying field is convolved with the complete Green's function, it is impossible to quantify the effect of the global observed SST pattern on the feedback. Since such a convolution would again lead to noisy timeseries of $N$ and $T$ and the subsequent need to compute a transient feedback via 30-year moving window regressions, Green's function approaches will likely also lead to strong statistical noise. Our results do not necessarily challenge the physically-consistent results form the Green's functions. Rather, it is possible that the spatial field of SSTs in an unforced \textit{piControl} simulation covaries in ways such that strong global fluctuations in the feedback are largely balanced out. It is therefore possible that a warming or cooling in a specific region in the Pacific does have the physically expected effect, however this effect is globally compensated by other radiative effects accumulating from elsewhere on the globe. In other words, the (univariate) signal from one specific region does not say anything about the simultaneous other signals coming from the multivariate pattern of surface warming.\newline

In summary, a persistent, strong and externally prescribed SST anomaly will produce a radiative response that is sufficient to infer the magnitude of the feedback parameter, consistent with decades of AMIP type experiments successfully used in model development and tuning (e.g., of cloud parameters) and Greens-function or pacemaker type experiments. Inferring changes in the feedback parameter from internally produced patterns, however, is hard or impossible.  

\section{Conclusions and Outlook}
Our work sheds light on two important limitations of using the previously established \textit{amip-piForcing} method to estimate the variability of Earth's feedback from observations. First, we have shown that the classical \textit{amip-piForcing} setup fails to capture the evolution of the true coupled system feedback, with a very low correlation between the two (Pearson corr. = 0.23). We also show that the \textit{amip-piForcing} runs exhibit a systematic negative bias in the feedback, in particular during the period 1940--2000. We next identify a major driver of this bias, namely the omission of prescribing atmospheric forcing (and in particular, aerosol forcing) in the \textit{amip-piForcing} setup. We then show that when prescribing the correct atmospheric forcing, the agreement between coupled and \textit{amip}-style feedback improves substantially (Pearson corr. = 0.68).
We therefore strongly recommend that any \textit{amip-piForcing}-type simulations ran to estimate the observed feedback parameter evolution be changed to \textit{amip-histForcing}-type runs, where the correct atmospheric forcing is also prescribed, as in the classical AMIP protocol \citep{eyring_overview_2016}. However, individual members still fail to reproduce the coupled feedback evolution, and there can be substantial differences in the sign of the feedback trends, despite identical SSTs, SIC, \textit{and} correctly prescribed atmospheric forcing (Fig.~\ref{fig:feedbacks_amipHistForcing}). In particular, the average ensemble bias towards an overall more stabilizing feedback in the \textit{amip}-style simulations remains, albeit reduced, even if the correct atmospheric forcing is prescribed. This result might imply that the true Earth feedback is in reality less stabilizing than inferred from \textit{amip}-style simulations. Our findings thus undermine the conclusions made on the basis of \textit{amip-piForcing} runs. Prescribing observed SSTs, SIC, and forcing to an atmospheric model appears to still be insufficient to constrain the coupled feedback evolution.  \newline

Our results are particularly worrying when considering that the experimental setup we provide is using model-internal SSTs and SIC as boundary conditions. In this very idealized case we would expect the \textit{amip}-setup---especially with prescribed atmospheric forcing as in the coupled setup---to reproduce the feedback variability of the coupled run, because everything happens in a "perfect model" world. In the \textit{amip-piForcing} runs with observations, however, SSTs and SIC with completely model-foreign mean-state patterns are fed to the model, trying to force the model to produce a pattern that it potentially internally would never do. The effect of this foreign SST and SIC prescription is not taken into account in our idealized setup and might bias the estimate even further. Our results are based on a single GCM and should therefore be validated with other models before confirming all our findings, but it is hard to imagine a method to produce a robust relationship across many models if it does not work in a single perfect model setup. We provide the experimental setup to do this validation and propose that it is done in order to assess the effectiveness of \textit{amip-piForcing} runs in future use. However, we point out that precisely because the \textit{amip-piForcing} setup fails even when prescribing model-internal SSTs and SIC to the same model, the leap of faith when prescribing observations to a model is even larger.\newline

Second, we show that estimating the feedback time series via 30-year moving window regressions introduces substantial statistical noise via the Yule-Slutzky effect.  This variability makes the feedback vary from values of -3 to above 0 $\mathrm{Wm^{-2}K^{-1}}$ in \textit{piControl} simulations. This low-frequency variability in the feedback could be purely statistical, as it occurs also in Monte-Carlo simulations with pure white noise, and also in the simulations with a covariance and autocorrelation structure of $N$ and $T$ as that in the fully-coupled \textit{piControl}. The timescale of this statistically-induced variability is 60--100 years. This finding alone implies that trends identified in the feedback timeseries on timescales shorter than 100 years could be purely statistical artifacts. This result helps explain how the \textit{amip}-style and coupled runs can show differences in the feedback despite identical SSTs, SIC, and atmospheric forcing. Our results are also in line with the findings of \citet{modak_better-constrained_2023}, who show that just changing the observational SST and SIC dataset used for forcing the \textit{amip-piForcing} simulations leads to substantially different feedback evolutions. We then demonstrate that periods of weakening or strengthening feedback cannot be statistically linked to specific trends in SST patterns in a fully-coupled 4000-year \textit{piControl} simulation. We find no significant SST trends emerging during periods of feedback strengthening or weakening, and virtually no agreement in SST trends between different periods of similar feedback change. We thus find no evidence of an unforced SST pattern effect in the CESM2 \textit{piControl} simulation.\newline 

Based on these results, we believe that trends detected based on \textit{amip-piForcing} runs over the past decades a) can not reproduce the corresponding feedback in the coupled system correctly, b) can not be causally linked to the evolving SST pattern, and c) may be entirely unphysical and a result of the amplification of statistical noise via the Yule-Slutzky effect. Since the statistically-induced variations peak at a timescale of 60--100 years, any variations detected in the feedback time series on periods shorter than this may be purely statistical and their attribution to SSTs, at least to our current knowledge and based on standard methods, is impossible. The problem is further complicated by the fact that signal-to-noise is comparatively weak in the recent period, as the real world is still far away from an idealized, strong forcing setup like a CO\textsubscript{2}-quadrupling. While the substantial differences between the CMIP6-simulated and the observed Pacific SST trends are undoubtedly real \citep[e.g.,][]{wills_systematic_2022,olonscheck_broad_2020}, the argument of a trend towards a more stabilizing, SST-driven feedback in recent decades based on \textit{amip-piForcing} simulations and found in a multitude of studies \citep[e.g.,][]{gregory_variation_2016,andrews_accounting_2018,zhou_analyzing_2017,dong_biased_2021,andrews_effect_2022,zhou_greater_2021} might be a) an effect of the misrepresentation of the coupled feedback in the atmosphere-only setup, b) an effect of the lack of prescribing atmospheric forcing, or c) a statistical effect. Based on the broad spectrum of strong limitations of both the current \textit{amip-piForcing} setup, but also on the feedback estimation via 30-year moving window regressions, we believe that this method provides little potential to constrain changes in the Earth's feedback over the historical period. 




\codedataavailability{Climate model simulations from CMIP6 used to generate the boundary conditions for our simulations are available at https://esgf-node.llnl.gov/projects/cmip6/ and were accessed through the ETH Zurich Next-Generation Archive \cite{brunner_eth_2020}.  We acknowledge the World Climate Research Programme, which, through its Working Group on Coupled Modelling, coordinated and promoted CMIP6. We thank the climate modeling groups for producing and making available their model output, the Earth System Grid Federation (ESGF) for archiving the data and providing access, and the multiple funding agencies who support CMIP6 and ESGF. Climate model simulations from the CESM2 single-forcing large ensemble used to generate the boundary conditions for our simulations are available via \url{https://gdex.ucar.edu/datasets/d651055/dataaccess/}. 

The code used for reproducing the analysis and all figures in the study will be made available at the time of publication at \url{https://doi.org/10.5281/zenodo.21937291} \citep{gyuleva_code_2026}. }



\appendix
\section{Methods for estimating the forcing $F$}\label{sec:app:meth_diagnosing_F}
\begin{figure}[h] 
	\centering
	\includegraphics[width=0.8\textwidth]{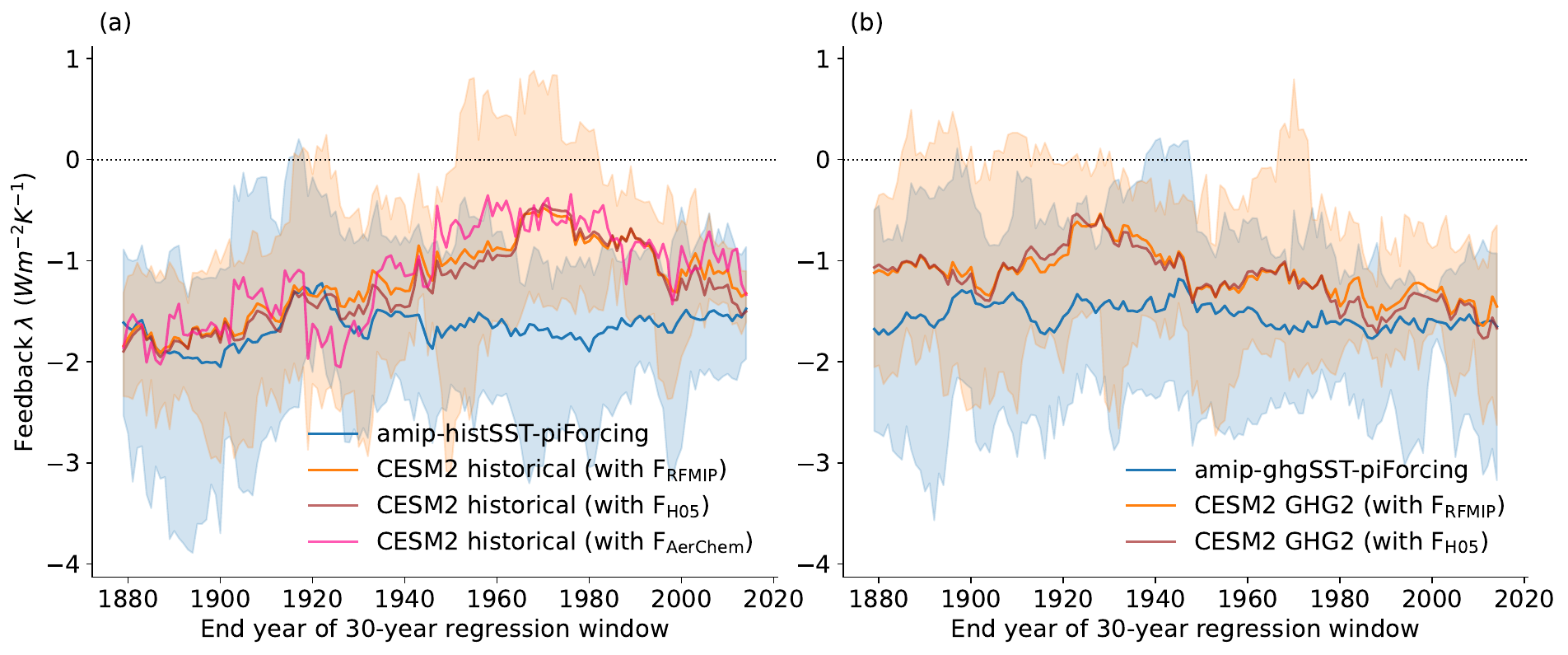}
\caption{Identical to Fig.~\ref{fig:feedback_amip_vs_cpld}, but with two alternative estimates of the feedback in the coupled runs calculated with the different $F$ estimates (see Appendix ~\ref{sec:app:meth_diagnosing_F}): the first correction adds the \citet{hansen_efficacy_2005} land correction to account for warming land masses (brown line). The second alternative $F$ estimate is based on the \citet{collins_aerchemmip_2017} method (pink line). The $F_{\mathrm{AerChem}}$ correction can only be estimated for the \textit{historical} setup, as the necessary runs for the GHG-only setup do not exist.}
	\label{fig:app:F-sensitivity}
\end{figure}

Estimating the feedback parameter requires estimates of the forcing $F$ in simulations with prescribed time-varying atmospheric forcing. We employ three different methods of estimating the forcing $F$ in order to test the sensitivity of our results to the forcing estimate: the RFMIP method \citep[][]{pincus_radiative_2016}, the RFMIP method with a correction proposed by \citet{hansen_efficacy_2005}, and the AerChemMIP method \citep{collins_aerchemmip_2017}. The three different $F$ estimates for the \textit{historical} forcing are shown in Fig.~\ref{fig:app:F-sensitivity}\newline

\begin{figure}[h] 
	\centering
	\includegraphics[width=0.8\textwidth]{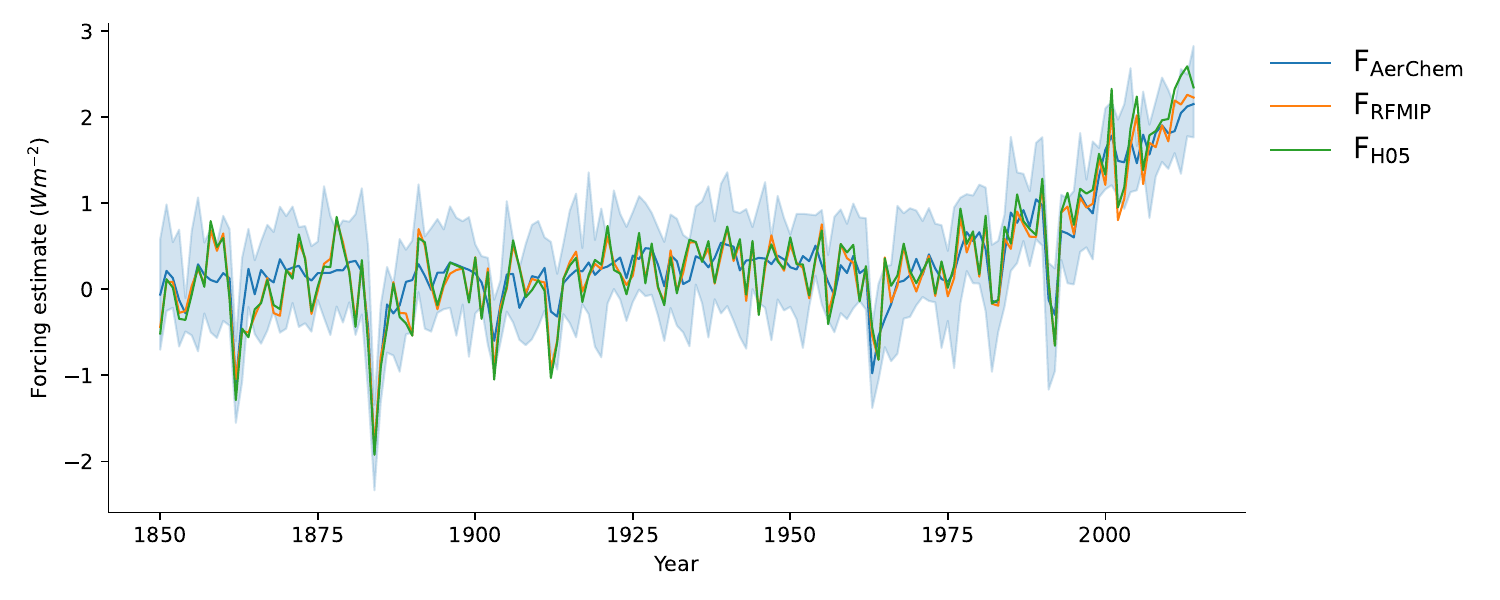}
\caption{Three different forcing estimates, obtained via the methods described above in Appendix \ref{sec:app:meth_diagnosing_F}. The blue shading shows the ensemble spread from the eleven members used to derive the $F_{\mathrm{AerChem}}$ estimate.}
	\label{fig:app:F-estimates}
\end{figure}

The RFMIP method for calculating the time-varying $F$ time series involves running fixed-SST experiments with time-varying forcing: the time-varying radiative forcing is estimated as the difference in net TOA imbalance between an atmosphere-only run with fixed climatological SSTs but time-varying atmospheric forcing, and a control run with climatological SSTs and fixed pre-industrial atmospheric forcing (\textit{piClim-control}). We perform a 75-year long \textit{piClim-control} simulation with monthly climatological SSTs taken from the coupled CESM2 \textit{piControl} simulation. For the scope of the forcing estimation 30 years would be sufficient \citep{forster_recommendations_2016}, but we use the \textit{piClim-control} also in our analysis of feedbacks (Sec.~\ref{sec:results_Variability}). We then also perform the \textit{piClim-histall} and \textit{piClim-ghg} simulations with the identical climatological SSTs as the \textit{piClim-control}, but time-varying atmospheric forcing (historical for the \textit{piClim-histall} and only historical GHG for the \textit{piClim-histghg}). We thereby obtain forcing time series $F_{\text{RFMIP}}$ for evolving historical and historical GHG-only forcing (orange line in Fig.~\ref{fig:app:F-estimates}). These forcing time series then correspond to our estimates of the forcing for the coupled \textit{historical} runs and the \textit{GHG2} runs, from which SST and SIC boundary conditions were obtained .\newline

The RFMIP method has the advantage of being computationally relatively simple and requiring only roughly 30 years of \textit{piClim-control} simulation, in addition to the time-varying $F$ experiment. In addition, due to the climatological prescribed SSTs, internal variability is largely suppressed, leading to a more accurate $F$ estimate \citep{forster_recommendations_2016}. However, the RFMIP method also exhibits two potential caveats. First, as only ocean temperatures are held fixed but the land is allowed to respond to the prescribed atmospheric forcing, part of the net TOA imbalance in the \textit{piClim-histall} and \textit{piClim-histghg} experiments is a response of the land to the forcing, rather than the forcing itself. The magnitude of this effect has only been quantified for forcing estimates of abrupt CO\textsubscript{2} quadrupling runs, where it has been shown to lead to an underestimation of the forcing by up to 14\% (although this result is based on a single GCM) \citep{andrews_effective_2021}. \newline

One way to correct for this approach is to employ the method proposed by \citet{hansen_efficacy_2005} (from now on denoted as H05). The net TOA response is corrected by an estimate of the response of the land given by $\lambda \Delta T_{\text{lnd}}$: 
\begin{equation}
    F_{\text{H05}} = F_{\text{RFMIP}} - \lambda \Delta T_{\text{lnd}}~~
\end{equation}
Since $\lambda$ is negative and the change in land temperature $\Delta T_{\text{lnd}}$ is positive in runs with historical or historical GHG-only atmospheric forcing, the H05 correction results in an overall upward correction of the forcing. \citet{andrews_effective_2021} find that this correction still underestimates the true $F$ estimate, yet their results are based on abrupt 4xCO\textsubscript{2} forcing, and for just one model. Due to the simplicity of the correction and its low computational cost, we decide to use the H05 correction to test the sensitivity of our results, despite its possibly limited accuracy (green line in Fig.~\ref{fig:app:F-estimates}). \newline

An alternative method to diagnose $F$ in transient runs is employed as part of the Aerosols and Chemistry Model Intercomparison Project (AerChemMIP) \citep{collins_aerchemmip_2017} and was first proposed by \citet{andrews_using_2014}. In this method, $F$ is diagnosed as the difference between an atmospheric simulation with time-evolving prescribed SSTs, SIC, and atmospheric forcing (such as a standard AMIP simulation \citep{eyring_overview_2016}, or the \textit{amip-histForcing-histSST} simulations), and a simulation with identical time-evolving SSTs and SIC, but no atmospheric forcing (such as \textit{amip-piForcing}-style runs). The benefit of this approach is that potential interactions of the forcing agents with the transient SST and SIC change are accounted for \citep{forster_recommendations_2016,collins_aerchemmip_2017}. However, the noise in the transient $F$ estimate due to variability is larger, calling for the need of an ensemble of simulations to derive a reliable estimate \citep{forster_recommendations_2016}. Here, we have such an ensemble: we compute the $F$ time series as the difference between the \textit{amip-histForcing-histSST} and the corresponding \textit{amip-piForcing-histSST} runs with pairwise identical SSTs. The resulting transient $F$ estimate, denoted as $F_{\text{AerChem}}$ is then taken as the ensemble mean of the eleven $F$ estimates from the ensemble (blue line in Fig.~\ref{fig:app:F-estimates}). 

\appendixfigures

\section{Robustness of spatial trends}
\begin{figure}[h] 
	\centering\includegraphics[width=0.8\textwidth]{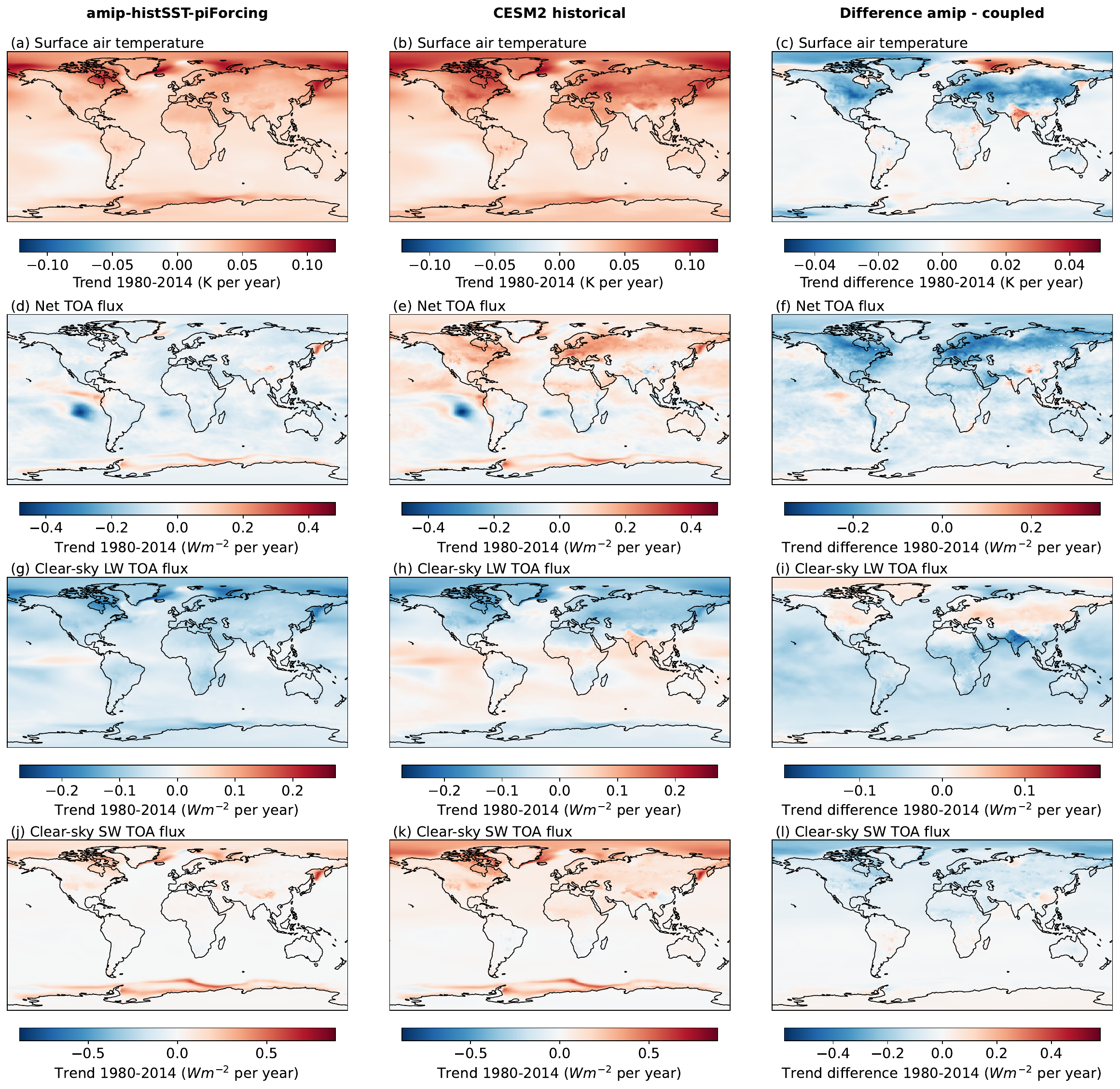}
\caption{As Fig.~\ref{fig:spatial_trends} of the main text, but with SW and LW flux trends computed on clear-sky fluxes.}
\label{fig:app:spatial_clearsky}
\end{figure}

\begin{figure}[h] 
	\centering
\includegraphics[width=0.8\textwidth]{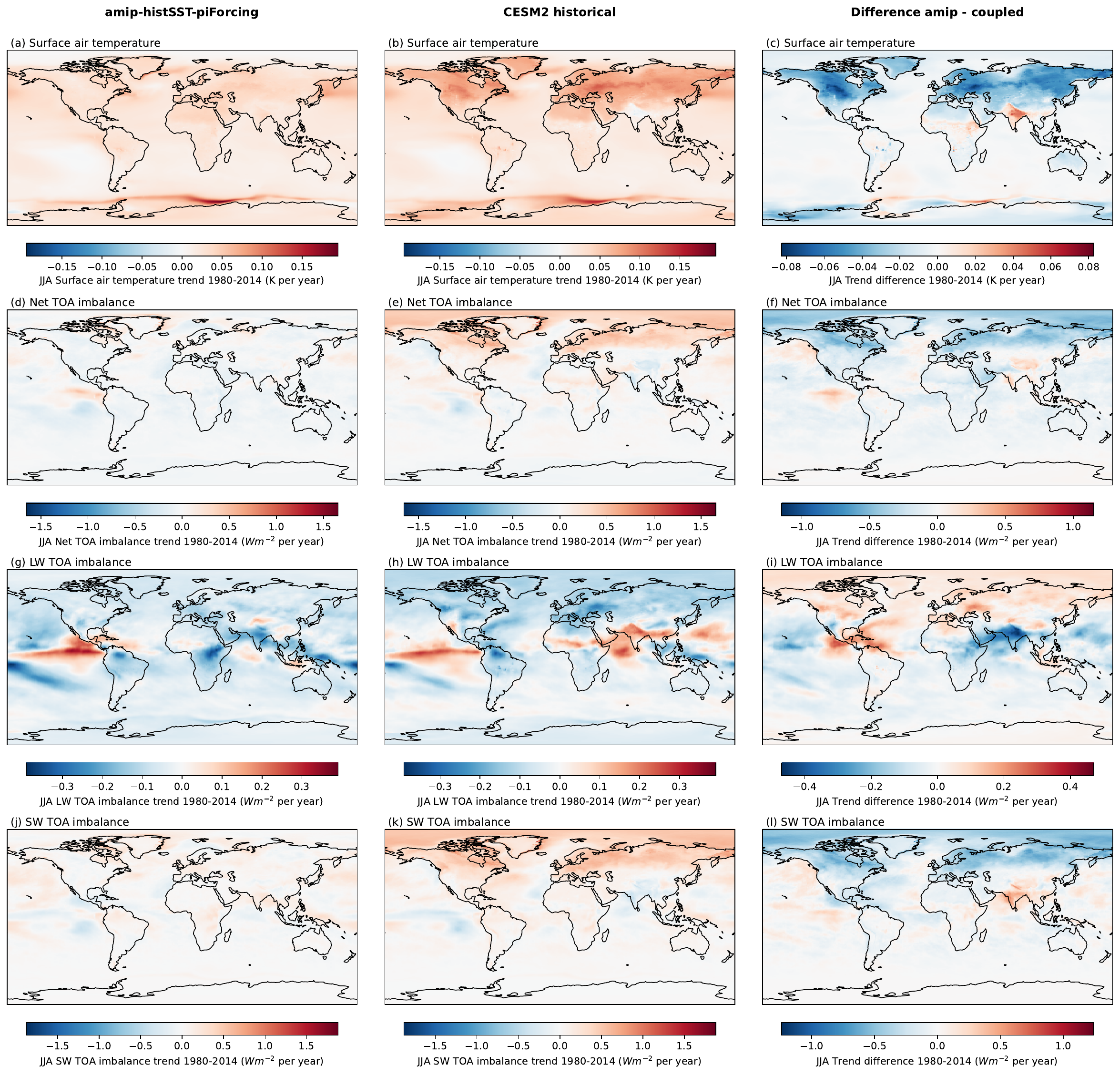}
\caption{As Fig.~\ref{fig:spatial_trends} of the main text, but with trends computed only for the summer season (June, July, August).}
	\label{fig:app:jja_spatial}
\end{figure}

\section{Perios of feedback weakening and strengthening in \textit{piControl-long}}
\begin{figure}[h] 
	\centering
\includegraphics[width=0.8\textwidth]{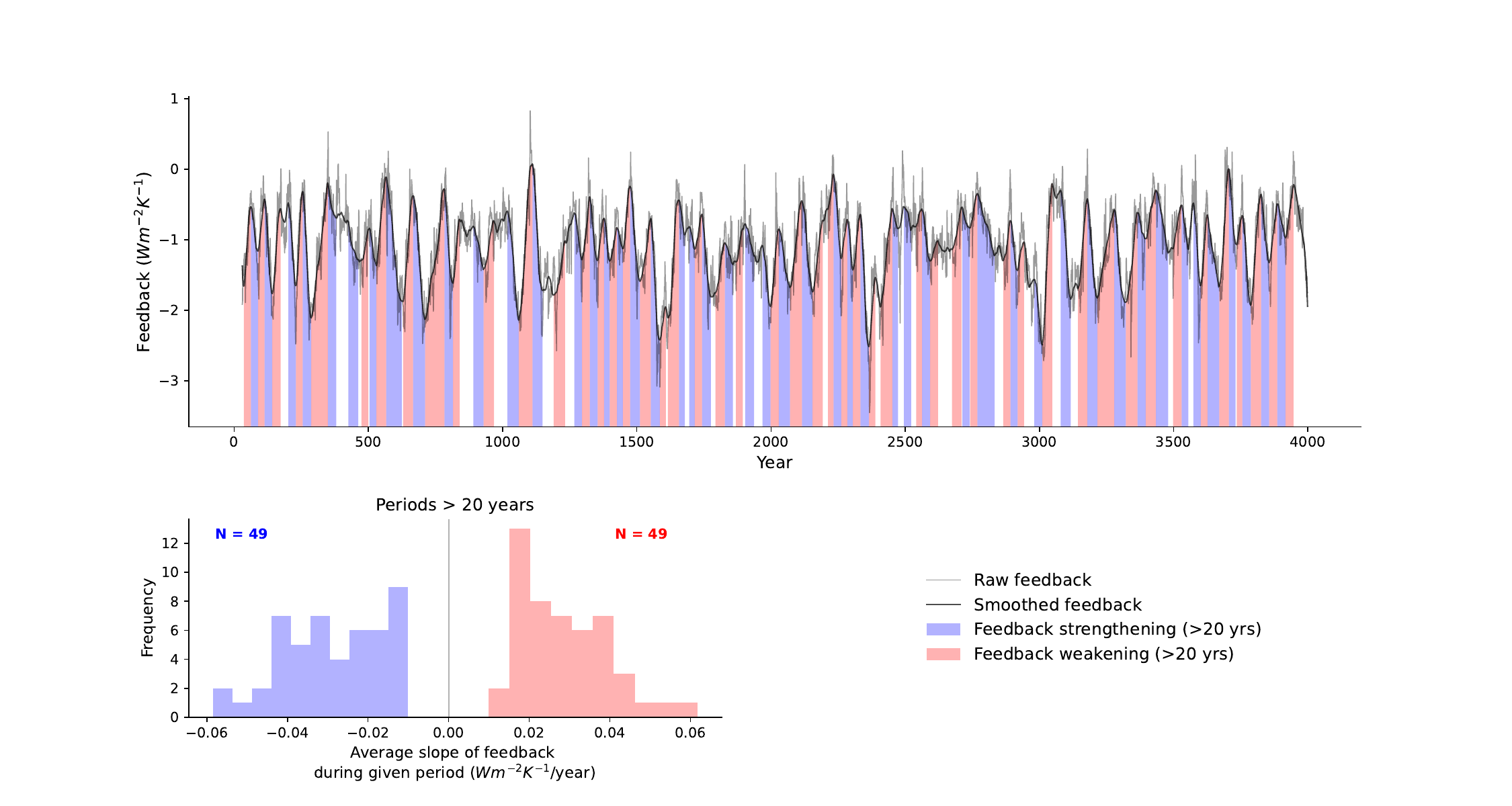}
\caption{(a): The raw (thin grey line) and smoothed (thick black line) feedback time series in the \textit{piControl-long} simulation. Red and blue periods indicate regions of at least 20 consecutive years identified as having an increasing (weakening) or decreasing (strengthening) feedback according to the methodology described in Sec.~\ref{sec:meth:bootstrap}. (b): Histogram of the average local slopes of the trends in the periods identified in (a).}
	\label{fig:app:composite_periods}
\end{figure}

\noappendix       




\appendixfigures  

\appendixtables   


\authorcontribution{GG devised the experiments, analyzed the data and wrote the manuscript. RN, RK, and SS developed ideas for the analysis. RK and SS provided supervision. RN, EF, RN, and SS contributed to the manuscript. UB ran the experiments and generated the data.} 

\competinginterests{The authors declare that they have no conflict of interest.} 


\begin{acknowledgements}
GG thanks Patricia Helpap, Florian Römer, Iris de Vries, and Konstantin Weber for the insightful discussions. GG, SS and RK acknowledge the project `Constraints on near-term warming projections via distributionally robust statistical and machine learning' (COPE; grant agreement C22-02, funded by the Swiss Data Science Center). SS acknowledges funding provided by the German Research Foundation through the Heinz Maier-Leibnitz Prize 2024 and the European Union's Horizon Europe programme via the project `Artificial Intelligence for Enhanced Representation of Processes and Extremes in Earth System Models' (AI4PEX; grant agreement 101137682). GG and RK acknowledge funding by the European Union's Horizon 2020 research and innovation programme under grant agreement No. 101003536 (ESM2025–Earth System Models for the Future). RN has received funding from the Swiss National Science Foundation (SNSF) via grant no. 216710.
\end{acknowledgements}

\bibliographystyle{copernicus}
\bibliography{references}

\end{document}